\documentclass[acmtog]{acmart}
\acmSubmissionID{1116}

\usepackage{enumitem}
\usepackage{graphicx}
\usepackage{ragged2e}
\usepackage{multirow}
\usepackage{multicol}
\usepackage{array}
\usepackage{listings}
\usepackage{xcolor}

\setcopyright{acmlicensed}
\acmJournal{TOG}
\acmYear{2025}
\acmVolume{44}
\acmNumber{4}
\acmArticle{}
\acmMonth{8}
\acmDOI{10.1145/3731209}

\newif\ifshownotes
\shownotestrue 

\ifdefined\MakeWithNotes
  \shownotestrue
\fi
\ifdefined\MakeWithoutNotes
  \shownotesfalse
\fi

\ifshownotes
  \newcommand{\colornote}[3]{{\color{#1}\bf{#2: #3}\normalfont}}
  \newcommand{\colornoteTwo}[3]{{\color{#1}\bf{#3}\normalfont}}
  \newcommand{\colornoteThree}[2]{{\color{#1}\bf{#2}\normalfont}}      
\else
  \newcommand{\colornote}[3]{}
  \newcommand{\colornoteTwo}[3]{}
  \newcommand{\colornoteThree}[2]{}      
\fi

\definecolor{darkgreen}{rgb}{0.0,0.65,0}

\newcommand{\camready}[1]{\textcolor{blue}{#1}}
\newcommand{\dslname}{MoVer}
\newcommand{\dslurl}{\url{https://mover-dsl.github.io}}

\renewcommand{\camready}[1]{#1}

\begin{document}

\title{\dslname{}: Motion Verification for Motion Graphics Animations}


\author{Jiaju Ma}
\affiliation{%
  \institution{Stanford University}
  \country{USA}}
\email{jiajuma@cs.stanford.edu}

\author{Maneesh Agrawala}
\affiliation{%
  \institution{Stanford University}
  \country{USA}}
\email{maneesh@cs.stanford.edu}

\renewcommand{\shortauthors}{Ma and Agrawala}

\begin{abstract}
While large vision-language models can generate motion graphics animations
from text prompts, they regularly fail to include all spatio-temporal properties described in the prompt. 
We introduce \dslname{}, a motion verification DSL based on
first-order logic that can check spatio-temporal properties of a
motion graphics animation.
We identify a general set of such properties that people commonly use
to describe animations (e.g., the direction and timing of motions,
the relative positioning of objects, etc.).
We implement these properties as predicates in \dslname{} and provide an execution
engine that can apply a \dslname{} program to any input
SVG-based motion graphics animation.
%
We then demonstrate how \dslname{} can be used in an LLM-based
synthesis and verification pipeline for iteratively refining motion graphics animations.
Given a text prompt, our pipeline synthesizes a motion graphics animation and a corresponding \dslname{} program. 
Executing the verification program on the animation yields a report of the predicates that failed
and the report can be automatically fed back to LLM to iteratively correct the animation.
%
To evaluate our pipeline, we build a synthetic dataset of
5600 text prompts paired with 
ground truth \dslname{} verification programs.
We find that while our LLM-based pipeline is able to automatically generate a
correct motion graphics animation for 58.8\% of the test prompts
without any iteration, this number raises to 93.6\% with up to 50
correction iterations.
\camready{%
Our code and dataset are at \dslurl{}.%
}%

\end{abstract}




\begin{teaserfigure}
    \centering
    \includegraphics[width=\textwidth]{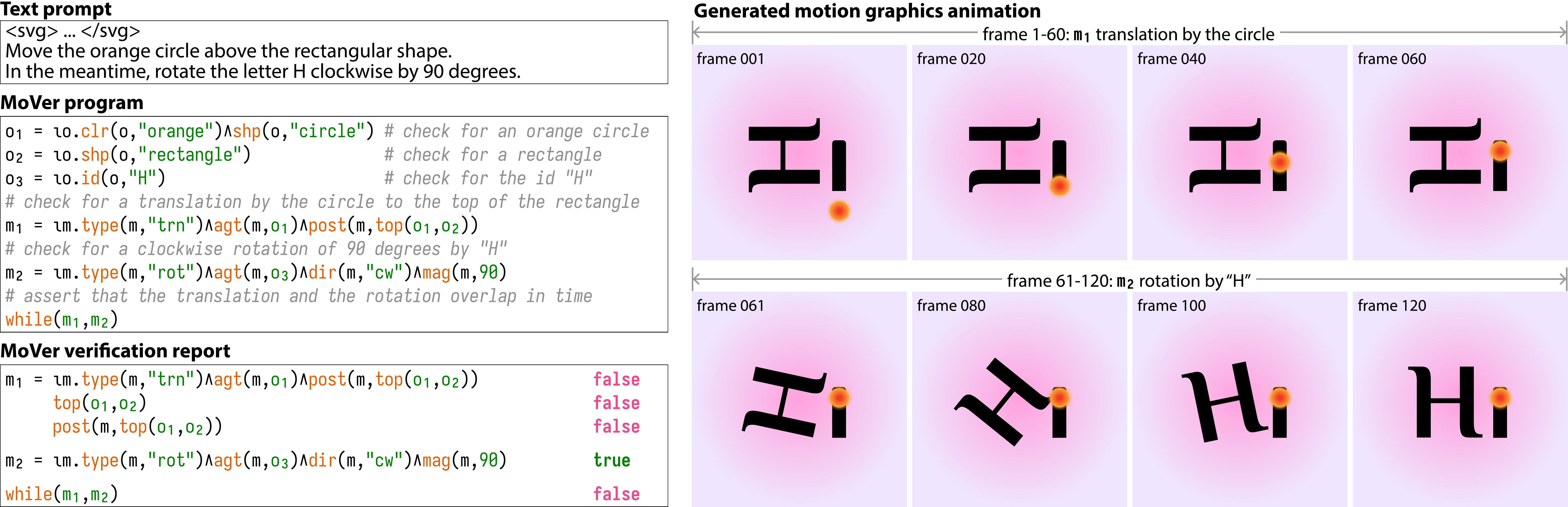}
     \vspace{-2em}
    \caption{
      We present \dslname{}, a motion verification DSL, and apply it in
      an LLM-based motion graphics synthesis pipeline. Given a text
      prompt (top left) describing a desired animation (with a static SVG
      scene for context) our pipeline synthesizes both (1)
      an SVG motion graphics animation (right) and (2) a corresponding
      \dslname{} verification program (middle left) that captures
      spatio-temporal properties mentioned in the prompt as
      first-order logic predicates.
      Executing the \dslname{} program on the animation
      produces a verification report (bottom left) that specifies
      which motions and predicates the animation includes (marked true) and
      which it does not (marked false).
      Here, {\tt m$_1$} is false because the circle motion does not
      end at the top of the rectangle.
      Motion {\tt m$_2$} is true because the H correctly rotates 90 degrees counterclockwise.
      The {\tt while()} predicate is false
      because {\tt  m$_1$} is invalid.
      Our pipeline can automatically feed the verification report back to the LLM-based animation
      synthesizer to iteratively correct the animation or users can examine the report to facilitate
      their debugging process.
    }
    \label{fig:teaser}
\end{teaserfigure}

\maketitle

\section{Introduction}
\label{sec:introduction}


Advances in large vision-language models have yielded
powerful tools for visual content creation via program synthesis.
%
These models can take a natural language text prompt as input and generate
program code in a domain specific language (e.g., SVG, CAD, etc.) that is then executed to
produce visual output including images\,\cite{gupta2023visprog}, vector
graphics\,\cite{xing2024llm4svg,polaczek2025neuralsvg},
3D scenes and CAD\,\cite{hu2025scenecraft,zhang2024scenelanguage,khan2024text2cadgeneration},
and animation\,\cite{gal2024breathing,tseng2024keyframer,liu2024logomotion}.
%
%
Yet, automatically verifying that the resulting program produces 
{\em correct} visual output---output that matches the
specification given in the prompt---remains a challenging task.

%
In some visual domains like image generation, researchers have
explored automatic verification via concept matching between the input
prompt and the output image\,\cite{Johnson_2017_ICCV,Hessel2021CLIPScoreAR,lee2023holisticevaluationtexttoimagemodels,hu2023tifa,JaeminCho2024}.
This approach relies on access to
concept detectors for both the natural language prompt
and for the output visual asset.
More importantly, the detected concepts must be relevant to the desired
properties of the generated content.
%
%
Verifying the prompt ``A horse sitting on an astronaut" might
require concept detectors for the entities ``horse'' and``astronaut''
and the spatial relationship ``sitting on''.

Here
we focus on such automated verification for 2D motion
graphics animations.
%
We first present {\bf \dslname{}}, a {\bf Mo}tion {\bf Ver}ification DSL, 
that is based on first-order logic
%
and can be used to check properties of the spatio-temporal trajectories of objects in an animation.
%
More specifically, in \dslname{}, checks are written as logic
statements constructed from predicates that verify the presence of
specific spatio-temporal properties such as the direction, the
magnitude, or the type (e.g., translation, rotation, scale) of motions
of an object. 
%
%
%
We identify relevant spatio-temporal concepts
based on prior work from cognitive psychology
that has investigated how people think about relationships between
objects and their motions in 
animation\,\cite{talmy1983language,talmy1975motion,tversky1998space}.
We show how to implement these concepts as  \dslname{} predicates 
using Allen's\,\shortcite{allen1983interval} temporal interval algebra
and its 2D spatial counterpart, the rectangle
algebra\,\cite{balbiani1998model,navarrete2013spatial}.

We then demonstrate how \dslname{} can be applied in an LLM-based 
synthesis and verification pipeline to iteratively produce motion graphics animation that accurately
match an input text prompt.
Given a text prompt specifying a desired animation (with a static SVG scene for context),
we use in-context LLM-based program
synthesis\,\cite{gupta2023visprog,surismenon2023vipergpt} to produce
(1) a motion graphics animation (as an SVG motion program) and (2) a corresponding \dslname{} 
verification program.
%
%
%
%
Executing the \dslname{} program on the corresponding
motion graphics animation results in a report pinpointing
the spatio-temporal predicates the animation successfully passes and fails (Figure~\ref{fig:teaser}).
Users can use the report to manually fix the animation program, or
automatically feed it back to our LLM-based pipeline for iterative correction.

To evaluate our synthesis and verification pipeline, we build a
synthetic text dataset of 5600 natural language text prompts describing
motion graphics animations, paired with ground truth \dslname{}
verification programs. We find that our LLM-based pipeline can
synthesize correct \dslname{} programs for 95.1\% of the test
prompts. We also find that using the automatic correction iteration capability of
our pipeline increases the number of correct animations from 58.8\%
(with no iterations) to 93.6\% (up to 50 iterations).
\camready{%
Our code and dataset are available at \dslurl{}.
}


\begin{table*}[t]
\footnotesize
\centering
\caption{The \dslname{} language provides logic predicates that checks for the fulfillment of various spatio-temporal properties of objects and motions in an input animation. Note that each object and motion attribute predicate has a corresponding  accessor form (e.g. {\tt get\_shp()}) that returns the value of that attribute.
}
\vspace{-1em}
\resizebox{\textwidth}{!}{%
\begin{tabular}{p{0.13\textwidth}p{0.2\textwidth}p{0.67\textwidth}}
\toprule
Name      & 
Syntax                   & 
Description \vspace{0.5mm}\\
\hline
variables & {\tt o$_i$}, {\tt m$_j$}          & 
Variables that represent objects and motions in the animation scene \vspace{0mm}\\
\hline
\multicolumn{3}{l}{\textit{Logical operators}} \\
\hline
$\forall$, $\exists$ &
\texttt{exists(var, expr)} &
\texttt{expr} is a logical statement, and \texttt{var} is an object or motion. Returns true if there is an assignment of \texttt{var} that satisfies \texttt{expr}. \\
$\iota$      & 
\texttt{iota(var, expr)}          & 
Returns an object or motion that satisfies \texttt{expr} and assign it to \texttt{var}. \\
A       & 
\texttt{all(var, expr)} & 
Return all objects or motions that satisfy \texttt{expr} and assign it to \texttt{var}.\\
$\neg$ & 
\texttt{not expr}& 
Compute the negation of \texttt{expr}. \\
$\land$, $\lor$   & 
\texttt{expr$_1$ and expr$_2$}   & 
Compute the conjunction or disjunction of \texttt{expr$_1$} and \texttt{expr$_2$} \vspace{0mm}\\
\hline
\multicolumn{3}{l}{\textit{Object attribute predicates}}\\
\hline
%
%
shape     & 
\texttt{shp(o$_i$, shp\_val)}     & 
Returns true if o$_i$ has the shape specified by \texttt{shp\_val}, and false otherwise. \\
color     & 
\texttt{clr(o$_i$, clr\_val)}     & 
Returns true if o$_i$ has the color specified by \texttt{clr\_val}, and false otherwise. \\
id     & 
\texttt{id(o$_i$, id\_val)}     & 
Returns true if o$_i$ has the id specified by \texttt{id\_val}, and false otherwise. \\
\hline
\multicolumn{3}{l}{\textit{Atomic motion attribute predicates}} \\
\hline
agent &
\texttt{agt(m$_j$, o$_i$)} &
Returns a vector $V$ of booleans (one per frame). If o$_i$ performs m$_j$ at frame $n$, then $v_n \in V$ is true, and false otherwise. \\ 
type      & 
\texttt{type(m$_j$, type\_val)}      & 
If m$_j$ has motion type of \texttt{type\_val} (either \texttt{trn}, \texttt{rot}, or \texttt{scl}) at frame $n$, then $v_n \in V$ is true, and false otherwise.    \\
direction & 
\texttt{dir(m$_j$, dir\_val)} & 
If m$_j$ is in the direction \texttt{dir\_val} at frame $n$, then $v_n \in V$ is true, and false otherwise.             \\
magnitude & 
\texttt{mag(m$_j$, mag\_val)} & 
If m$_j$ reaches the magnitude \texttt{mag\_val} over a sequence of frames $m,...,n$, then $v_m,....,v_n \in V$ are true, and false otherwise.\\
%
origin    & 
\texttt{orig(m$_j$, orig\_val)}   & 
If m$_j$ is performed around the origin \texttt{orig\_val} at frame $n$, then $v_n \in V$ is true, and false otherwise.    \\
duration  & 
\texttt{dur(m$_j$, dur\_val)}  & 
If m$_j$ lasts the duration \texttt{dur\_val} over the sequence of frames, $m,...,n$, then $v_m,....,v_n \in V$ are true, and false otherwise.
\\
post &
\texttt{post(m$_j$, expr)} &
If \texttt{expr} evaluates to true on the last frame of {\tt m$_j$}, then $v_n \in V$ is true for all frames of {\tt m$_j$}, and false otherwise.\\
\hline
\multicolumn{3}{l}{\textit{Spatial and temporal relative predicates}} \\
\hline
temporal relative &
  \texttt{<tmp\_rel>(m$_j$, m$_j$)} &
  \texttt{<tmp\_rel>} is one of the 13 Allen's interval algebra predicates or aggregations of them (Figure~\ref{fig:allen_rectangle} top). Returns true if the time intervals of m$_j$ and m$_j$ satisfy \texttt{<tmp\_rel>}.
      \dslname{} includes three aggregated temporal predicates: \texttt{before()}, \texttt{while()}, and \texttt{after()}.

  \\
spatial relative &
  \texttt{<spt\_rel>(o$_i$, o$_j$)} &
  \texttt{<spt\_rel>} is one of the 169 rectangle algebra predicates or aggregations of them (Figure~\ref{fig:allen_rectangle} bottom). Returns true if the spatial positions of o$_i$ and o$_j$ satisfy \texttt{<spt\_rel>}.
  \dslname{} implements ten aggregated predicates: \texttt{top()}, \texttt{bottom()}, \texttt{left()}, \texttt{right()}, \texttt{border()}, \texttt{intersect()}, \texttt{top\_border()}, \texttt{bottom\_border()}, \texttt{left\_border()}, and \texttt{right\_border()}.
\\
\bottomrule
\\
\end{tabular}%
}
\vspace{-2em}
\label{tab:predicates}
\end{table*}

\section{Related Work}
\label{sec:related}

\noindent
{\bf \em Text-to-animation.}
Generating animations from a natural language prompt has been a long
standing challenge in computer graphics\,\cite{bouali2023review}.
Early text-to-animation methods often used rule-based natural language
processing techniques such as regular expression parsing, co-reference
resolution, and sentence simplification
to map the prompt to a semantic representation (e.g. a scene graph) that is rendered
into an
animation\,\cite{badler1999,aakerberg2003carsim,ma2006automatic,marti2018cardinal,zhang2019generating}.
%
%
While effective for specialized domains such as posing virtual
humans\,\cite{badler1999} or simulating car
accidents\,\cite{aakerberg2003carsim}, such methods typically
require many special case rules to handle the ambiguity of general
natural language. 

In the current era of large generative AI models, researchers have explored end-to-end methods for directly
generating animation from text prompts without passing
through a semantic
representation\,\cite{guo2023animatediff,guo2023sparsectrl,xing2024tooncrafter,chen2024videocrafter2,gal2024breathing}.
%
%
While these end-to-end methods can generate impressive animations, they
regularly fail to capture all of the desired spatio-temporal
properties mentioned in the text prompt.
%
%
%
In contrast, we focus on automatically verifying such properties in synthesized
motion graphics animations.

%
%
%


\vspace{0.5em}
\noindent
{\bf \em LLM-based program synthesis.} 
Our work builds on recent techniques that use LLMs for 
program synthesis via in-context learning\,\cite{gupta2023visprog,surismenon2023vipergpt}.
These methods typically provide language documentation with code examples
(pairs of text prompts with corresponding programs) as part of the
system prompt context.
The approach has been applied in a variety of visual synthesis domains
including 
images\,\cite{gupta2023visprog}, vector
graphics\,\cite{xing2024llm4svg}, 3D
scenes\,\cite{zhang2024scenelanguage}, CAD\,\cite{jones2025solver}, and human
motion animation\,\cite{Goel_2024}.
%
Closest to our work are Keyframer\,\cite{tseng2024keyframer} and
LogoMotion\,\cite{liu2024logomotion} which apply this LLM-based
program synthesis approach to the problem of generating 2D vector
animations.
While all these methods generate a human-editable program
representation for the visual content, none of them provide automated
verification capable of checking that the resulting visuals match
the prompt specification.
%
It remains up to the user to manually identify mismatches and fix the
resulting program or to update their prompt and try again.
Our work aims to close this gap in the context of motion graphics 
by providing automated verification and thereby allowing
the LLM to iteratively correct the animation without user
intervention.

\vspace{0.5em}
\noindent
{\bf \em Visual verification.}
%
%
Our work is inspired by recent efforts that use visual question
answering (VQA) models to check that properties mentioned in an input
prompt appear in diffusion generated
images\,\cite{Johnson_2017_ICCV,hu2023tifa,JaeminCho2024}.
For example, these methods convert the image generation prompt ``A
tree behind a motorcycle'' into a series of natural language
questions like ``Does the image contain a tree?'' and ``Is the tree behind
the motorcycle?'' and use a VQA model to report the answer to each
one.
The answers can then be converted into a score or examined
individually by users.
\camready{Design2Code\,\cite{si2024design2code}, on the other hand, proposes a set of automatic metrics for evaluating webpage generation quality.}
%
%
%
%
DreamSync\,\cite{sun2024dreamsync} incorporates the VQA-based
review of visual outputs into an iterative optimization loop where the output from
VQA model is automatically fed back to the image generating model.
\camready{%
SceneCraft\,\cite{hu2025scenecraft} and LLMR\,\cite{torre2024llmr} apply this optimization approach
in the context of 3D and VR scene generation respectively, using LLM-based question answering.%
}
%
%
Our work similarly provides an iterative optimization loop for motion
graphics animation, but, instead of a VQA model, it uses an LLM to convert the prompt into a first-order
logic based verification language.

%
%
%

\vspace{0.5em}
\noindent
{\bf \em Mapping language into formal logical specifications.}
One approach to representing the semantic meaning of natural language
expressions is to map the expression into formal logical
specifications.  For example, Artzi et
al.\,\shortcite{artzi2013mapping} show how to map English language
navigational phrases such as 
``move to the chair'' to a
corresponding logical form \texttt{$\lambda$a.move(a)$\land$to(a,$\iota$x.chair(x))}.
Researchers have explored this mapping approach in the
context of instructing robotic
manipulators\,\cite{wang2023programmatically}, 
VQA\,\cite{hsu2024left} and guidance for text-to-image
generation\,\cite{sueyoshi2024predicated}.
These methods often use specialized semantic parsing
techniques\,\cite{kamath2018survey} to convert natural language into
the formal logic.
While our \dslname{} verification language is also based on
formal, first-order logic, we use the program synthesis capabilities of
an LLM to map prompts to \dslname{} predicates. We show that our
LLM-based approach handles a wider variety of inputs than a rule-based
semantic parser.


\section{\dslname{}: A Motion Verification DSL}
\label{sec:moverDSL}

\dslname{} is a domain specific language for verifying that certain
spatio-temporal concepts appear in a motion graphics animation.
%
We first consider the design of the \dslname{} language (Section~\ref{sec:design}) and then describe
how we execute a \dslname{} verification program (Section~\ref{sec:execution}).

\subsection{\dslname{} Language Design}
\label{sec:design}

We designed \dslname{} based on prior work from cognitive psychology
that has analyzed how people think about and describe spatio-temporal properties and relationships in animated
scenes\,\cite{talmy1983language,talmy1975motion,tversky1998space}.
%
%
%
%
This work shows that linguistic descriptions of animation
schematize the information to emphasize only the most salient aspects
of the motions of each object.
People break complex trajectories into basic motions; they
individually describe the translation, rotation, and scaling of an
object even when all three occur simultaneously. 
People also commonly use relative frames of reference rather than absolute frames,
e.g. saying ``Move the circle above the square after rotating the
triangle.'' instead of ``Move the circle to (100,100) starting at second 5.0''.

In \dslname{}, spatio-temporal concepts are expressed as
first-order logic predicates (boolean functions)
over variables representing objects {\tt o$_i$} $\in \mathbf{O}$ and
their motions {\tt m$_j$} $\in \mathbf{M}$
(Table~\ref{tab:predicates}).
\dslname{} treats seven attributes as 
{\em atomic} descriptions of a motion {\tt m$_j$}:
\begin{enumerate}[leftmargin=0.5cm]
    \item {\tt \bf agt } The agent object {\tt o$_i$} performing motion {\tt m$_j$}.
    \item {\tt \bf type}  The motion type of {\tt m$_j$}: ``translation,'' ``rotation,'' or ``scale''. 
    \item {\tt \bf dir } The direction of motion {\tt m$_j$}.
    \item {\tt \bf mag } The magnitude or distance of motion {\tt m$_j$}.
    \item {\tt \bf orig} The origin of motion {\tt m$_j$} when {\tt type} is rotation or scale.
    \item {\tt \bf dur } The duration of motion {\tt m$_j$}.
    \item {\tt \bf post} A postcondition the end state of motion {\tt m$_j$} must satisfy.
\end{enumerate}
\dslname{} includes predicates that can check the values for each of these
attributes. For example, the predicate {\tt dir(m$_j$, [1,0])} returns
true for each frame where motion {\tt m$_j$} is along the direction vector
$[1,0]$ (i.e., rightwards).
We can combine individual predicates into a complete logic statement
to check whether an animation contains a motion {\tt m} with a set of \textit{atomic} properties as
\begin{multline}
\exists \texttt{m.type(m,"trn")}\land\texttt{dir(m,[0,1])}\land\texttt{mag(m,100)}\\ \land\texttt{agt(m,$\iota$o.clr(o,"black")}\land\texttt{shp(o,"square"))}.
\end{multline}
This statement returns true for frames containing a motion where a black
square-shaped object is translating upwards by 100 px.

\begin{figure}[t]
  \centering
  \includegraphics[width=\linewidth]{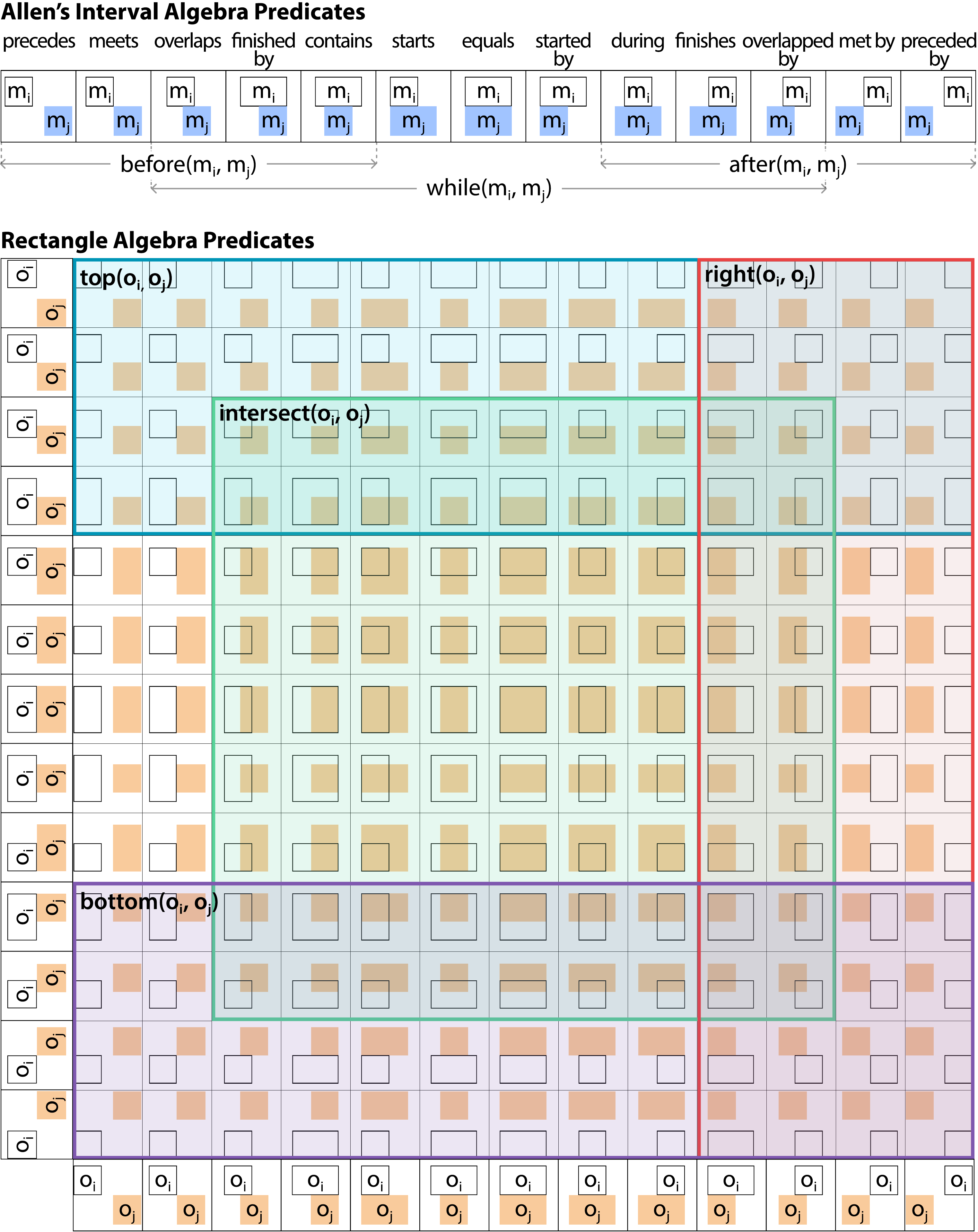}
  \vspace{-2em}
  \caption{
      (top) The 13 low-level temporally relative predicates in Allen's interval algebra and the three aggregated predicates (disjunctions of the low level predicates) in \dslname: \texttt{before()}, \texttt{while()}, and \texttt{after()}. 
      (bottom) The 169 spatially relative predicates in the rectangle algebra and four of the aggregated predicates in \dslname{}. 
      \dslname{} implements 10 spatially relative aggregated predicates: \texttt{top()}, \texttt{bottom()}, \texttt{left()}, \texttt{right()}, \texttt{border()}, \texttt{intersect()}, \texttt{top\_border()}, \texttt{bottom\_border()}, \texttt{left\_border()}, and \texttt{right\_border()} as sub-regions of this table.
  }
    \vspace{-2em}
  \label{fig:allen_rectangle}
\end{figure}

\dslname{} also includes predicates that can check relative
spatio-temporal relationships between motions.
These predicates are based on Allen's~\shortcite{allen1983interval}
temporal interval algebra and its 2D analog, the rectangle
algebra~\cite{balbiani1998model,navarrete2013spatial}.
These algebras provide a formal representation for 13 temporally
relative relationships and 169 spatially relative relationships (Figure~\ref{fig:allen_rectangle}) and
\dslname{} includes predicates that can check for all of them.
In addition, \dslname{} aggregates certain relative predicates together with
disjunction to form higher-level relative predicates to better accommodate for natural
language's inherent ambiguity.
For example, the higher-level aggregated temporal predicate
\texttt{while(m$_i$, m$_j$)} (Figure~\ref{fig:allen_rectangle} top)
returns true
for each frame where both {\tt m$_i$} and {\tt m$_j$}  occur at the same time and aggregates
nine lower-level Allen algebra predicates.
%
%
Similarly, the aggregated spatial predicate {\tt top(o$_i$,o$_j$)}
returns true as long as the top edge of object {\tt o$_i$} is above
the top edge of {\tt o$_j$} and aggregates 52 lower-level rectangle
algebra predicates.

\subsection{\dslname{} Execution Engine}
\label{sec:execution}
%
%

We have implemented an execution engine for \dslname{} that can check
whether a motion graphics animation is consistent with a \dslname{}
program.
It computes the set of frames that evaluate to false with respect to
each logical statement and predicate in the verification program and
outputs a verification report for each such expression
(an example report is Figure~\ref{fig:teaser} bottom left).


The \dslname{} execution engine operates on a generic, frame-level SVG
representation of motion graphics animations.
Specifically, each foreground object is represented as an SVG {\tt
  <shape>} , {\tt <g>}, or {\tt <image>} node with motion transforms
specified per-frame.
This SVG motion program representation is similar to that used by Zhang et
al.\,\shortcite{zhang2023motion}, and many high-level animation APIs,
including CSS~\cite{cssAnimation}, anime.js~\cite{animeJs}, and GSAP~\cite{gsap},
support rendering an animation into this type of
per-frame SVG representation.

Given such an SVG motion program, our execution engine first builds a
2D {\em animation matrix} data structure where each row represents an object and each
column represents a frame of the animation.  In each cell $(i,j)$ of
the animation matrix we store all atomic motion attributes (Section~\ref{sec:design}),
except {\tt post}, for object {\tt o$_i$} at frame $j$.
%
%
We compute these motion attributes by first factoring the per-frame
transformation matrix of each object into a translation, rotation, and
scale~\cite{thomas1991} and then depending on the type of motion,
computing a direction, magnitude, and origin.
We separately store static information for each object such as its color,
shape, and bounding box in object local coordinates.

With these data structures in place, we implement our DSL as follows.
%
%
We execute each logical predicate in the verification program on each
cell of the animation matrix. We mark the
cell as true if the
predicate is satisfied and false otherwise. Thus each predicate
produces a boolean matrix with the same shape as the animation matrix.
For example, executing the predicate {\tt type(m,"rot")} sets
matrix cell $(i,j)$ to true if object {\tt o$_i$} has a non-zero rotation
angle in frame $j$. Similarly, executing the predicate {\tt shp(o$_i$,"square")}
sets the entire matrix row $(i,*)$ to true if
object {\tt o$_i$} is square-shaped. 
A predicate expression (i.e., {\tt expr}) combines the results of
multiple predicates by applying logical boolean operators (e.g., $\land$,
$\lor$, $\lnot$) element-wise on the predicate matrices.
We can check logical quantifiers (e.g., $\exists$, $\forall$) by aggregating
the boolean matrices.
Finally, the $\iota$ operator returns an object $o_i$ or motion $m_j$
for which a logical expression {\tt expr} holds. We compute the
expression on the animation matrix, identify a row for which the
expression holds for at least one frame, and return the row index (if an object
is requested) or the row itself (if a motion is requested).
%

While many predicates primarily involve looking up information in the
animation and object data structures, temporal and spatial predicates
based on the interval and rectangle algebras are more complex.
These require comparisons between the starting and ending frames of
the motions or between the relative positions and overlaps of the
bounding boxes of the objects (Figure~\ref{fig:teaser}).
An interval tree representation~\cite{cormen2022introduction} of the
motions (2D intervals per-frame for spatial predicates, frame
discretized 1D intervals for temporal predicates) enables efficient
execution of these predicates.




\section{LLM-Based Motion Program Synthesis and Verification}
\label{sec:llmsynth}
We leverage \dslname{} within an motion graphics synthesis and
verification pipeline.
\begin{figure}[h]
  \centering
  \vspace{-1em}
  \includegraphics[width=\linewidth]{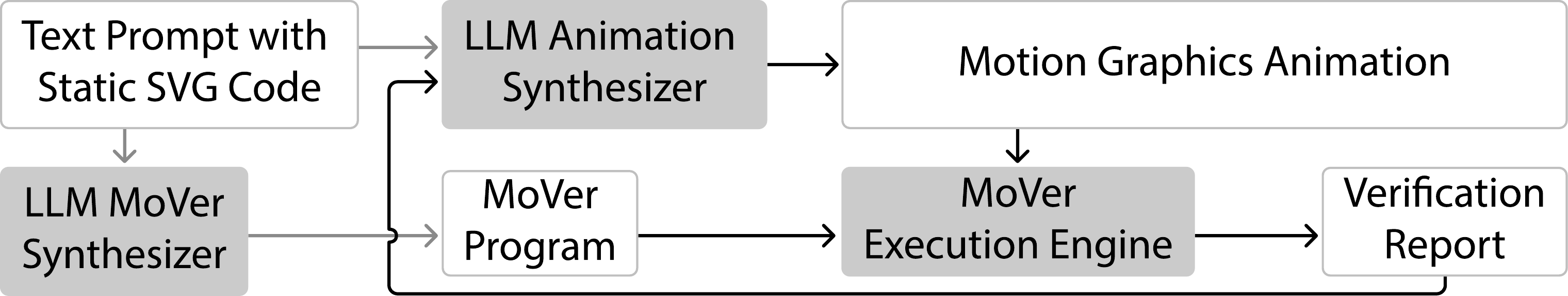}
  \vspace{-2em}
  \label{fig:pipeline}
\end{figure}%
It takes as input a text prompt describing the desired animation along with a static SVG that sets the scene context for the prompt.%
\begin{figure*}[t]
    \centering
    \includegraphics[width=\linewidth]{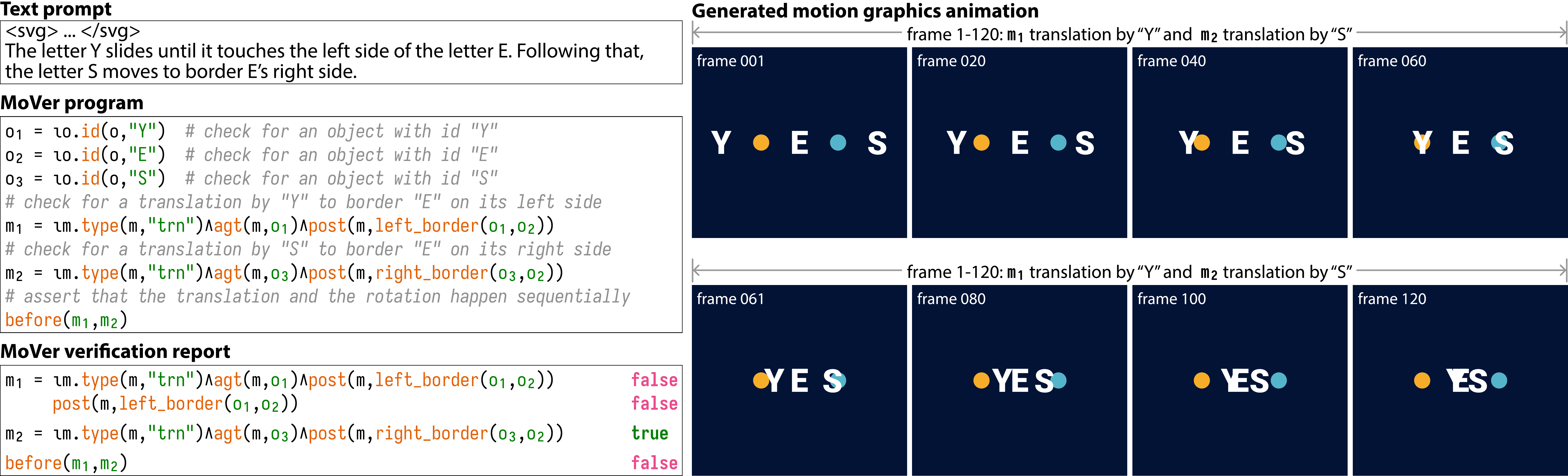}
    \vspace{-2em}
    \caption{The input text prompt (top left) aims to move the Y and
      the S so that they touch the E on either side.
      The verification report (bottom left)
      shows that, although the Y does border the left side of E on
      frame 80 and \texttt{left\_border()} evaluates to true, it
      continues its rightward motion \texttt{m}$_1$ until frame 120,
      and so the \texttt{post()} predicate fails.  
      Although \texttt{m}$_2$, the motion of S, successfully borders the right side of E, the \texttt{before()} predicate fails because \texttt{m}$_1$ is false.
      }
    \label{fig:verification}
\end{figure*}
\begin{figure*}[t]
    \centering
    \includegraphics[width=\linewidth]{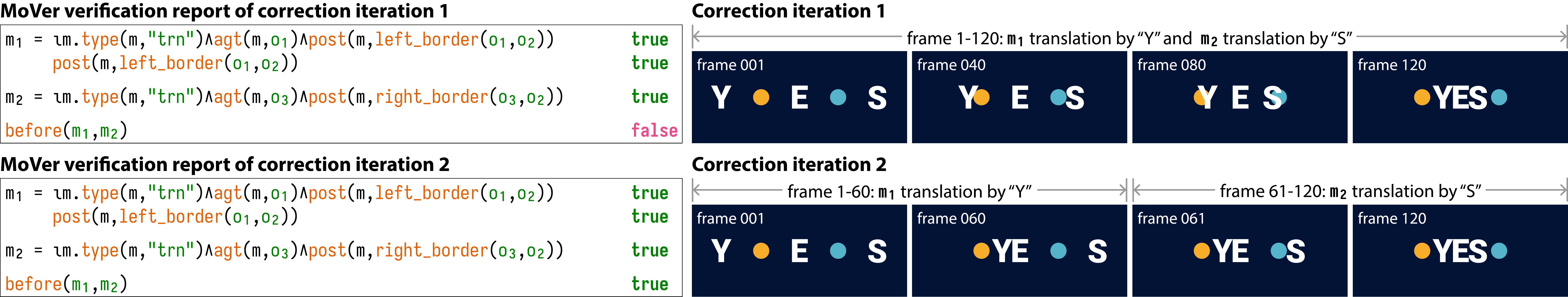}
    \vspace{-2em}
    \caption{Our pipeline automatically feeds the verification report
      (Figure~\ref{fig:verification}) back to the LLM animation
      synthesizer to correct the animation. On the first 
      iteration, the motion of Y correctly ends adjacent to the left
      side of E at frame 120. But the motion incorrectly overlaps in time with
      the motion of S. On the second correction iteration, the
      sequencing of the motions is correctly updated so that Y moves first
      (frames 1-60) before S (frames 61-120).  }
    \label{fig:iteration}
\end{figure*}
%
%
It uses LLM-based program synthesis to produce two programs;
(1) the {\em animation synthesizer} produces a motion graphics animation program written in a high-level
animation API (GSAP\,\cite{gsap} in our implementation)
and (2) the {\em \dslname{} synthesizer} generates a corresponding \dslname{} verification program.
In both cases we follow the LLM-based in-context learning approach
for program synthesis\,\cite{gupta2023visprog,
  surismenon2023vipergpt}, where we prompt GPT-4o\,\cite{hurst2024gpt}
with annotated API or DSL documentation and examples of text prompts
paired with corresponding output programs
\camready{%
(Appendix~\ref{sec:sysprompts} gives the prompts used in our implementation).%
}%

We render the resulting motion graphics animation into the generic
per-frame SVG motion program representation (Section~\ref{sec:execution}).
Then we execute the \dslname{} program on
the SVG motion program to check that spatio-temporal properties mentioned
in the prompt appear in the animation.
%
%
The output is a verification report
specifying which spatio-temporal properties appear in the animation and which do not.
%
If the verification report indicates a failure, our pipeline can automatically send the 
report back to the animation synthesizer
and request it to update the motion graphics
animation, thereby supporting iterative correction.
Note that when the feedback is automatically sent back to the
LLM, we include the documentation of our \dslname{} DSL in the prompt so it
can interpret the error report
\camready{%
(Appendix~\ref{sec:sysprompts:correction}).%
}
%
In addition, a human user can examine the verification report and either
manually fix the motion program, or filter the feedback sent to the
LLM. 
%
On each iteration, the LLM animation synthesizer has access to all
previous animations it has generated and the corresponding
verification reports in its chat history to help it make corrections.
%


\begin{figure*}[t]
    \centering
    \includegraphics[width=\linewidth]{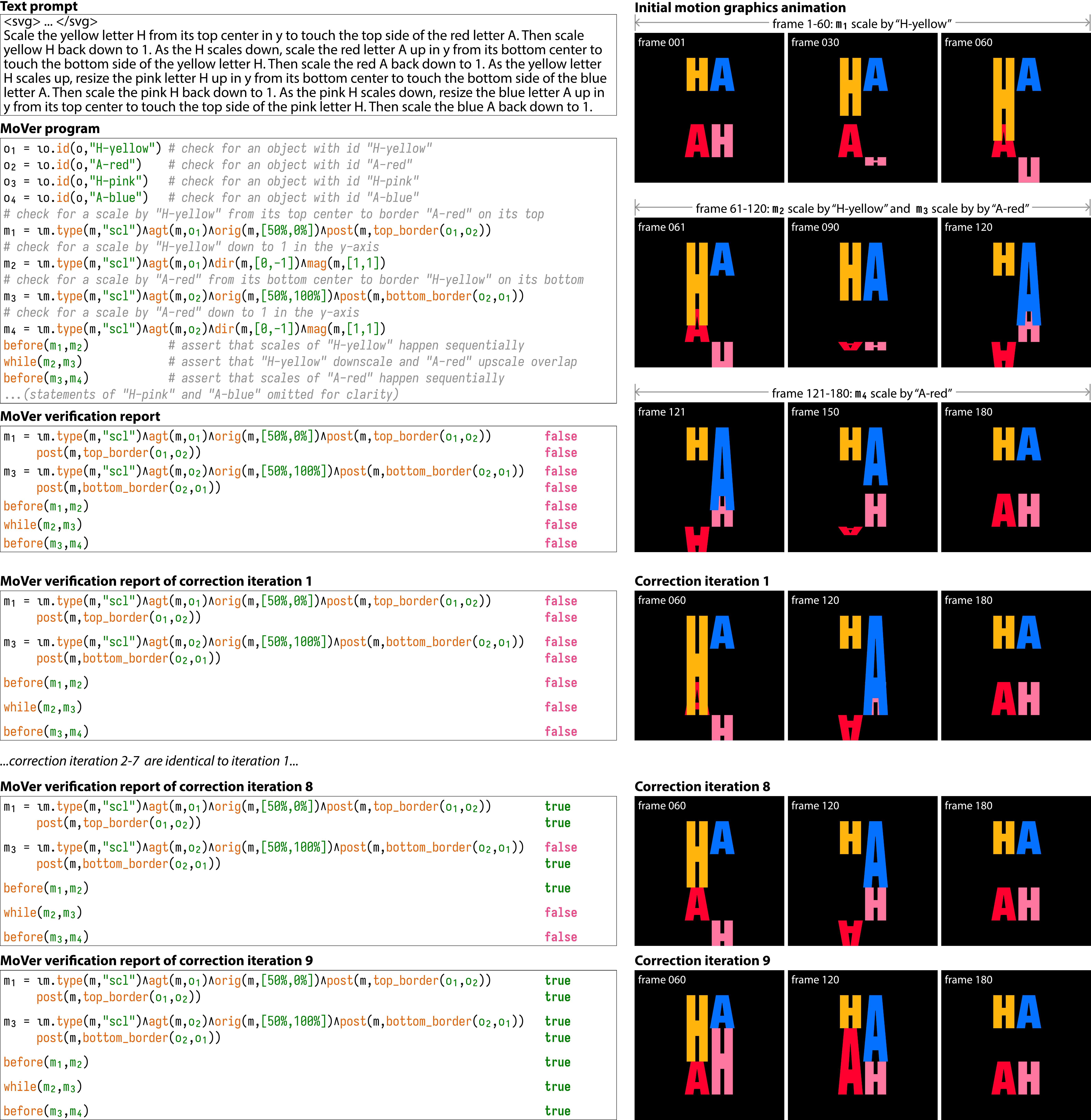}
    \caption{
        \camready{%
        The input text prompt describes an animation in which the letter H and the letter A are asked to maintain contact as they scale up and down, as shown in the frames under Correction Iteration 9 (please refer to \dslurl{} for the animations in action).
        However, the initial animation produced by the LLM animation synthesizer fails to satisfy any of the \texttt{border()} relations as indicated by the verification report.
        In the first correction iteration, the letters H and A are stretched to border the bottom edges of the letters below them, instead of top edges (frame 60 and 120).
        From there, the synthesizer becomes stuck in a loop, producing the same animation until the eighth correction iteration, where it has corrected the scaling of the yellow H and blue A.
        In the ninth iteration, it eventually changes the sign of the scale factor of the two bottom letters, passing all verification checks.%
        }%
    }
    \label{fig:haha}
\end{figure*}

\begin{figure*}[t]
    \centering
    \includegraphics[width=\linewidth]{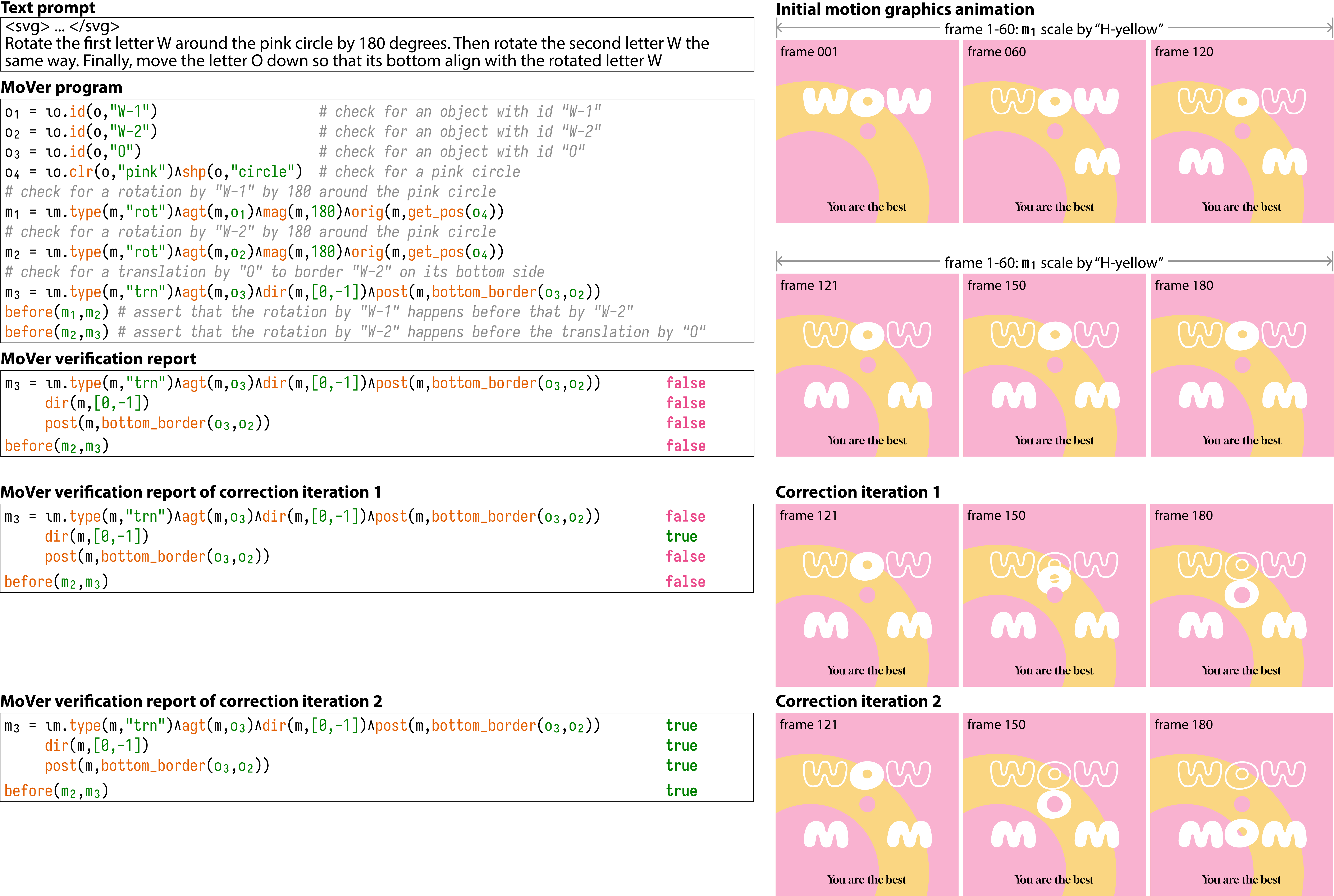}
    \vspace{-2em}
    \caption{
        \camready{%
        The input text prompt asks the two letter W's to be rotated around the small pink circle and the letter O to be lowered so that the transformed letters form the word ``mom'' (animations can be found at \dslurl{}).
        In the initial animation, the W's are rotated correctly, but the letter O does not move at all, rendering both \texttt{dir()} and \texttt{post()} to false.
        Note that \texttt{bottom\_border()} is true because the bottom edge of O aligns with those of the W's at the beginning of the animation (frame 1).
        In the first correction iteration, the O moves towards the bottom but does not do so enough to align with the W's.
        In the second iteration, the LLM synthesizer correctly computes the translation distance for O, producing an animation that says ``wow mom'' at the end (frame 180).%
        }%
    }
    \vspace{-0.75em}
    \label{fig:wowmom}
\end{figure*}

\section{Results}
\label{sec:results}

Figures~\ref{fig:teaser}, \ref{fig:verification}, \ref{fig:haha} and~\ref{fig:wowmom} show
examples of motion graphics animations generated with our
synthesis and verification pipeline
\camready{%
(more examples can be found at \dslurl{}).%
}%

%
%

In Figure~\ref{fig:teaser}, the prompt asks for an upward
translation of the orange circle above the rectangle and a 90-degree
rotation of the letter H with both motions overlapping in
time. The intention is that the animation should end after the shapes
form the word ``Hi.''
While the LLM-generated \dslname{} program captures the spatio-temporal
properties described in the prompt, 
the LLM-generated animation 
does not.
As shown in the animation frames and found by the verification
report, there is no valid motion \texttt{m}$_1$,
where the circle
translates to the top of the rectangle, and so both \texttt{top()}
and \texttt{post()} are false.
A valid \texttt{m}$_2$ is found as the letter H was
rotated by 90 degrees clockwise.
However, since \texttt{m}$_1$ fails, the temporal
predicate \texttt{while()} also fails since one of its input motions is
invalid.
%
%

In Figure~\ref{fig:verification}, the prompt describes a
series of motions that aim to move the letters Y and S adjacent to the letter 
E.
The \dslname{} program captures these desired properties and
the verification report
indicates that \texttt{m}$_1$, the motion of the letter Y, 
is false because \texttt{post()} failed,
even though its input predicate \texttt{left\_border()}
evaluates to true.
This is because 
at frame 80, the letter Y does border
E on its left side.  However, because the Y continues its translational motion
\texttt{m}$_1$ until frame 120, the \texttt{post()} predicate fails.
Since \texttt{m}$_1$ is invalid, the temporal
predicate \texttt{before()} also fails.
Figure~\ref{fig:iteration} shows the next two automatic correction iterations of our
pipeline for the example in
Figure~\ref{fig:verification}.
In correction iteration 1, the verifier reports that the letter Y
now properly borders the letter E at the end of the translation motion \texttt{m}$_1$.  Although
both \texttt{m}$_1$ and \texttt{m}$_2$ are valid, the temporal predicate
\texttt{before()} still fails because the motions do not
occur sequentially.
In the second iteration, the animation synthesizer fixes this
problem, and every statement and predicate in the \dslname{}
program evaluates to true, confirming that the final animation
correctly matches the input prompt.
Figures~\ref{fig:haha} and~\ref{fig:wowmom} similarly walk through examples that
require multiple correction iterations.
\camready{Please refer to their captions for more details.}


\section{Evaluation}
\label{sec:evaluation}

\vspace{0.5em}
\noindent
{\bf \em Test dataset.}
To evaluate our LLM-based synthesis and verification pipeline, we require
a test dataset consisting of text prompts describing motion
graphics animations and their corresponding ground truth \dslname{}
programs.
Since no such dataset is readily available, we synthetically
build one using a template-based approach in combination with an
LLM.
\camready{
Appendix~\ref{sec:prompt_generation} details our dataset construction.%
}
We divide the test dataset into four categories based on the types of \dslname{} predicates they exercise.
\begin{enumerate}[leftmargin=0.5cm]
    \item {\bf \em Single atomic motion.}  Each prompt describes
      one atomic motion with up to seven of its attributes as in Table~\ref{tab:predicates} (e.g., ``Expand the square by 2.'').
      
    \item {\bf \em Spatially relative motion.}  Each prompt
      describes one atomic motion that is relative to the spatial
      position of another object to exercise predicates from the
      rectangle algebra (Figure~\ref{fig:allen_rectangle} bottom) (e.g., ``Move the square next to the circle.'').
    
    \item {\bf \em Temporally relative motions.}  Each prompt describes 
      multiple atomic motions with relative temporal sequencing relations to
      exercise predicates from the Allen temporal interval algebra (Figure~\ref{fig:allen_rectangle} top). (e.g., ``Move the square,
      while scaling the circle'').
    
    \item {\bf \em  Spatio-temporally relative motions}: Each prompt
      describes multiple atomic motions with both spatially and temporally relative
      relationships (e.g., ``Move the circle next to the square, while it rotates'').
\end{enumerate}
\noindent
Our test dataset includes a total of 5600 pairs of prompts and \dslname{} programs (1200 single
atomic, 1400 spatially relative, 1200 temporally
relative, and 1800 spatio-temporally relative).
%
%
\camready{We release this dataset to promote further research at \dslurl{}.}

\begin{table}[t]
\footnotesize
\centering
\caption{
Success rate for our LLM-based \dslname{} program synthesizer and a baseline semantic parser.
We determine the success rate by first computing the string difference
between the LLM synthesized \dslname{} program and the corresponding ground truth
program. We count a success only when there is no difference.
We note that requiring a perfect string match is a 
stringent test for success as the underlying semantics of
the two programs may match even without an exact string match.
\vspace{-1em} 
%
}
\resizebox{\linewidth}{!}{%
\begin{tabular}{@{}lccccc@{}}
\toprule
\multirow[c]{2.7}{*}{\begin{tabular}[c]{@{}l@{}}\dslname{} program\\synthesis method\end{tabular}} & \multicolumn{5}{c}{Prompt Type}                     \\ \cline{2-6}
                                 & \begin{tabular}[c]{@{}c@{}}Single\\atomic\end{tabular} & \begin{tabular}[c]{@{}c@{}}Spatially\\relative\end{tabular} & \begin{tabular}[c]{@{}c@{}}Temporally\\relative\end{tabular} & \begin{tabular}[c]{@{}c@{}}Spatio-\\temporally\end{tabular}  & \multicolumn{1}{|c@{}}{Overall} \\
\hline
Ours (LLM-based)                 & \textbf{96.5\%}   & \textbf{97.1\%}            & \textbf{97.7\%}            & \textbf{90.9\%} & \multicolumn{1}{|c@{}}{\textbf{95.1\%}}    \\
Semantic Parser                & 87.1\%   & 86.3\%            & 92.4\%            & 76.8\% & \multicolumn{1}{|c@{}}{84.7\%}    \\
\bottomrule
\end{tabular}%
}
\label{tab:eval_parsing}
\end{table}

\vspace{0.5em}
\noindent {\bf \em Accuracy of LLM-based \dslname{} synthesizer.}
Table~\ref{tab:eval_parsing} reports the success rate for our
LLM-based \dslname{} synthesizer across our test dataset.
Overall we find that it 
generates correct \dslname{} programs for 95.1\% of the test prompts.
It is especially successful for prompts across the first three categories
(above 96\%), but less so for prompts in the spatio-temporally
relative category at 90.9\%.
Analyzing failure cases, we find that the LLM tends to
confuse relative spatial and temporal concepts.  For example, with a
prompt of the form ``The square translates. After that, the
circle scales,'' the LLM \dslname{} synthesizer sometimes produces the predicate
\texttt{after(m$_1$,m$_2$)} instead of \texttt{before(m$_1$,m$_2$)}.
%
%
There are also cases where the LLM confuses relative
positioning with absolute directions.  For the prompt ``Move the
square to the right of the circle,'' the LLM can produce
\texttt{dir(m,[1,0])} and \texttt{right(o$_1$,o$_2$)} instead of just
the latter, confusing rightward motion with the right-side
positioning.
%
Nevertheless, the overall success rate of the LLM \dslname{}
synthesizer suggests that it can produce correct programs
across a variety of prompts.

\begin{table}[t]
\centering
\caption{%
\camready{%
Performance of our iterative LLM-based synthesis and verification
pipeline (a)
and two variations of which we remove (b) and simplify (c) the verification reports. 
For our dataset of 5600 prompts, we compute the number that requires 0 correction
iterations (pass@0), the number that requires 1 to 49 iterations
(pass@1+), and the number that fails after 49 iterations.
We report the average number of iterations and their min-max ranges for the pass@1+ prompts.
Note that failed prompts are excluded from these metrics.
The slight differences in pass@0 between pipelines are the results of using a temperature of 1.0.
We use a seed of 1 to mitigate the inherent randomness of LLM generation.%
}%
}
\vspace{-1em} 
\resizebox{\linewidth}{!}{%
\begin{tabular}{@{} lrrrrr @{}}

\multicolumn{4}{@{}l}{\textbf{(a) Our LLM-based pipeline}} \\
\toprule
\multirow{2}{*}{\begin{tabular}[c]{@{}c@{}}\end{tabular}} & \multicolumn{5}{c}{Prompt Type}                     \\ \cline{1-6}
                                 & \begin{tabular}[c]{@{}c@{}}Single\\atomic\end{tabular} & \begin{tabular}[c]{@{}c@{}}Spatially\\relative\end{tabular} & \begin{tabular}[c]{@{}c@{}}Temporally\\relative\end{tabular} & \begin{tabular}[c]{@{}c@{}}Spatio-\\temporally\end{tabular}  & \multicolumn{1}{|c}{Overall} \\
\hline
pass@0                 & 1060 (88.3\%) & 530 (37.9\%)   & 1062 (88.5\%)   & 640 (35.6\%) & \multicolumn{1}{|r@{}}{3292 (58.8\%)}    \\
pass@1+                & 136 (11.3\%) & 718 (51.3\%)    & 134 (11.2\%)    & 962 (53.4\%) & \multicolumn{1}{|r@{}}{1950 (34.8\%)}    \\
fail              & 4 (0.3\%)    & 152 (10.9\%)    & 4 (0.3\%)       & 198 (11.0\%) & \multicolumn{1}{|r@{}}{358 (6.4\%)}    \\
\cline{1-6}
avg. iters.             &  1.1 (1-5)   & 6.2 (1-34)       & 1.8 (1-12)      & 6.7 (1-38) & \multicolumn{1}{|r@{}}{5.8 (1-38)}    \\
\bottomrule
\\

\multicolumn{4}{@{}l}{\textbf{(b) No feedback pipeline}} \\
\toprule
\multirow{2}{*}{\begin{tabular}[c]{@{}c@{}}\end{tabular}} & \multicolumn{5}{c}{Prompt Type}                     \\ \cline{1-6}
& \begin{tabular}[c]{@{}c@{}}Single\\atomic\end{tabular} & \begin{tabular}[c]{@{}c@{}}Spatially\\relative\end{tabular} & \begin{tabular}[c]{@{}c@{}}Temporally\\relative\end{tabular} & \begin{tabular}[c]{@{}c@{}}Spatio-\\temporally\end{tabular}  & \multicolumn{1}{|c}{Overall} \\
\hline
pass@0                 & 1062 (88.5\%) & 528 (37.7\%)   & 1064 (88.7\%)   & 646 (35.9\%) & \multicolumn{1}{|r@{}}{3300 (58.9\%)}    \\
pass@1+                & 42 (3.5\%)    & 314 (22.4\%)   & 118 (9.8\%)     & 200 (11.1\%) & \multicolumn{1}{|r@{}}{674 (12.1\%)}    \\
fail                   & 96 (8.0\%)    & 558 (39.9\%)   & 18 (1.5\%)      & 954 (53.0\%) & \multicolumn{1}{|r@{}}{1626 (29.0\%)}    \\
\cline{1-6}
avg. iters.            & 7.0 (1-32)   & 3.8 (1-35)      & 1.6 (1-10)      & 2.8 (1-38) & \multicolumn{1}{|r@{}}{3.3 (1-38)}    \\
\bottomrule
\\

\multicolumn{4}{@{}l}{\textbf{(c) Minimal feedback pipeline}} \\
\toprule
\multirow{2}{*}{\begin{tabular}[c]{@{}c@{}}\end{tabular}} & \multicolumn{5}{c}{Prompt Type}                     \\ \cline{1-6}
                                 & \begin{tabular}[c]{@{}c@{}}Single\\atomic\end{tabular} & \begin{tabular}[c]{@{}c@{}}Spatially\\relative\end{tabular} & \begin{tabular}[c]{@{}c@{}}Temporally\\relative\end{tabular} & \begin{tabular}[c]{@{}c@{}}Spatio-\\temporally\end{tabular}  & \multicolumn{1}{|c}{Overall} \\
\hline
pass@0                 & 1061 (88.4\%) & 532 (38.0\%)   & 1059 (88.3\%)  & 636 (35.3\%) & \multicolumn{1}{|r@{}}{3288 (58.7\%)}    \\
pass@1+                & 129 (10.8\%)  & 418 (29.9\%)   & 133 (11.0\%)   & 394 (21.9\%) & \multicolumn{1}{|r@{}}{1074 (19.2\%)}    \\
fail                   & 10 (0.8\%)    & 450 (32.1\%)   & 8 (0.7\%)      & 770 (42.8\%) & \multicolumn{1}{|r@{}}{1238 (22.1\%)}    \\
\cline{1-6}
avg. iters.            & 2.6 (1-21)    & 3.6 (1-37)     & 1.6 (1-8)      & 6.7 (1-37)   & \multicolumn{1}{|r@{}}{4.4 (1-37)}    \\
\bottomrule
\end{tabular}%
} 
%
\label{tab:eval_pipeline}
\end{table}




To better understand the generalization capabilities of our LLM-based
\dslname{} synthesizer, we implemented a classical rule-based semantic
parser~\cite{kamath2018survey} for converting a text prompt into a
\dslname{} program. This parser serves as a baseline.
\camready{%
Appendix~\ref{sec:semanticParser} describes the parser in detail.%
}
%
%
Table~\ref{tab:eval_parsing} shows that, although the semantic
parser achieves a high overall success rate of 83.5\%, it is
not as strong as our LLM-based approach.
Examining the unsuccessful prompts, we find that the 
semantic parser
has trouble generalizing to the variety of sentence structures found in 
our test set prompts.
For example, to describe an upward motion of
an object, a prompt might say ``add an upward translation to ...'' or ``propel
the object to climb ...''  The former embeds the motion type
as a noun, while the latter uses multiple verbs to express the
same atomic motion. Our semantic parser does not contain the rules necessary to handle the former
case and produces two motions for the latter.
In general, rule-based methods are brittle as it is difficult to cover all
the cases that can occur in natural language. 
%
%

Yet, even though our LLM-based approach generalizes well to a wide
variety of sentence structures and expressions, we find that it can
fail unexpectedly.  For example, it can randomly fail to include a
\texttt{mag()} predicate explicitly mentioned in the prompt (e.g., ``Rotate the square by 90 degrees'').
It might also output the wrong parameters for predicates,
such as flipping the sign of a directional parameter for \texttt{dir()}
or using \texttt{[1,0]} for an upward translation instead of
\texttt{[0,1]}.
There is no clear pattern to such failures and we also find that many 
similar prompts can produce correct results.
%
%
%

\begin{figure*}[t]
  \centering
    \includegraphics[width=\linewidth]{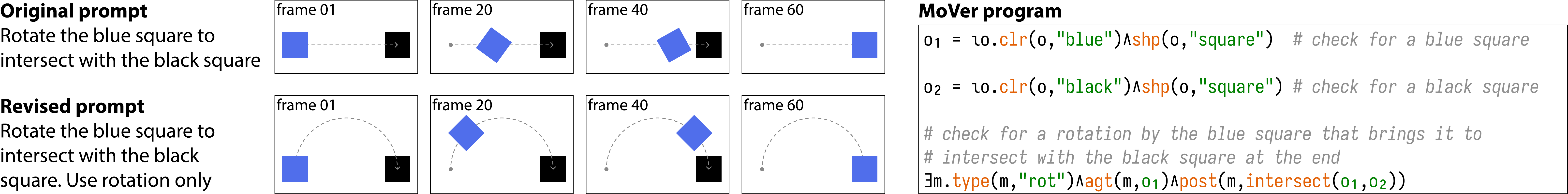}
    \vspace{-2em}
    \caption{\dslname{} can help user's identify ambiguous
      prompts. Our pipeline correctly generates a \dslname{} program
      reflecting the user's original prompt and it generates a motion that successfully
      passes verification. However, the user's intention was for the
      blue square to make an arcing motion rather
      than rotating about its center and simultaneously
      translating. Understanding why the animation passed verification
      but did not meet their intentions, the user revised the prompt to
      ask for a rotation-only motion and produced the desired result.
    }
    \vspace{-1em}
    \label{fig:discussion}
\end{figure*}

\vspace{0.5em}
\noindent {\bf \em Effectiveness of LLM-based synthesis and verification pipeline.}
Table~\ref{tab:eval_pipeline}a evaluates the effectiveness of
our fully automatic LLM-based motion graphics synthesis and verification
pipeline.
%
It shows that with 0 correction iterations the LLM-based animation
synthesizer produces a motion graphics animation that successfully passes all the
checks in the ground-truth \dslname{} program for 58.8\% of our test
prompts (pass@0).  It requires one or more correction
iterations to successfully generate another 34.8\% of the test prompts
(pass@1+), and it fails to pass the ground-truth \dslname{} program
after 49 correction iterations for 6.4\% of the test prompts (fail).
We also see that for the pass@1+ test prompts, the average number of
iterations is 5.8 with a (min-max) range of 1-38.
These numbers suggest that the automatic verification
and correction iterations in our pipeline (pass@0 and pass@1+)
significantly improves the number of correct animations produced,
compared to using only a single forward pass of LLM-based
animation synthesis alone (pass@0).

We observe that correction iterations are especially effective
for the spatially and spatio-temporally relative prompts, as the
iterations fix 51.3\% and 53.4\% (pass@1+) of these cases
respectively.
Examining the test prompts in these categories, we find that
initially our LLM-based animation synthesizer often neglects the
height or width of objects when translating one to a position relative
to another, but such errors are often corrected within a couple of iterations.
%
Examining the test prompts in these categories, we find that
initially the LLM-based animation synthesizer
has trouble interpreting
certain relative spatial concepts. 
For example, for the test prompt ``Translate the black square to touch
the blue circle,'' the synthesizer might initially move the square to
fully overlap the circle, rather than moving it to the border of the circle.
Errors like these can often be corrected within a
couple of iteration as the feedback from the verification report allows
the LLM to fix the problem.
%

%
Test prompts that require more correction iterations often ask for a
rotation or scale motion to position one object relative to another.
For example, with the prompt ``The blue circle is rotated to lie
on the top side of the black square'',  the LLM-based
animation synthesizer sometimes defaults to using translation despite
the explicit specification otherwise in the prompt.
%
It insists on using a translation for a few iterations, but after the
verification report marks the motion as failing, it eventually
corrects the motion to use a rotation.
%

\camready{%
To better understand the role of the \dslname{} verification reports in providing feedback for the correction iterations, we compare them to two alternative forms of feedback: (1)  
a \textit{no feedback} report where we simply ask the LLM animation synthesizer to generate another animation without any feedback (Table~\ref{tab:eval_pipeline}b), and (2) a \textit{minimal feedback} report where we only indicate whether an animation program passes or fails verification as a whole and do not include which predicates pass or fail (Table~\ref{tab:eval_pipeline}c).
%
%
Overall, the results show that providing no or minimal feedback makes the LLM animation synthesizer less effective at iterative correction, as it fixes fewer prompts (pass@1+ decreases from 34.8\% with \textit{our pipeline} to 12.1\% with \textit{no feedback} and 19.2\% with \textit{minimal feedback}) and increases the number of prompts that fail (fail increases from 6.4\% with \textit{our pipeline} to 29.0\% with \textit{no feedback} and 22.1\%  with \textit{minimal feedback}) across all prompt types. 
%
Spatially and spatio-temporally relative prompts produce the largest numbers of failures, with \textit{no feedback} (39.9\% and 53.0\%) performing worse than \textit{minimal feedback} (32.1\% and 42.8\%).
We note that the average iterations metrics decrease from 5.8 with \textit{our pipeline} to 2.3 with \textit{no feedback} and 4.4 with \textit{minimal feedback} because we exclude failed prompts from these numbers.
}

While the numbers in Table~\ref{tab:eval_pipeline} are based on
assuming the pipeline has access to ground truth \dslname{} programs,
in practice our pipeline generates \dslname{} program
using an LLM.
As shown in Table~\ref{tab:eval_parsing}, our pipeline generates
incorrect \dslname{} programs for 272 (4.9\%) test prompts. Running
these prompts through our full pipeline with incorrect \dslname{}
programs, we find 70 (25.7\%) pass@0, 20 (7.4\%) fail, 134 (49.3\%)
pass@1+ and these require an average of 6.7 (1-32) iterations, while
48 (17.6\%) generate \dslname{} programs that fail to execute.
In other words, running our full pipeline on these prompts would
falsely report that it produced correct animations for 204 (75\%) of
the prompts and that it produced incorrect animations for the
remaining 68 (25\%)  prompts.

\camready{%
Finally, we have also run the evaluation in Table~\ref{tab:eval_pipeline}a with LLMs other than GPT-4o (i.e. o3-mini, Gemini 2.0 Flash, Llama 3.1 8B and Llama 3.3 70B).
We find that our automatic verification and iterative corrections can increase the number of correct animations produced by LLMs regardless of their number of parameters.
Please refer to Appendix~\ref{sec:additional_test} and \dslurl{} for more details.
}

\vspace{0.5em}
\noindent
{\bf \em Discussion.}
\camready{%
Our experiment on the LLM-based \dslname{} synthesizer (Table~\ref{tab:eval_parsing}) shows that it is relatively good at converting natural language prompts of motions into \dslname{} verification programs (95.1\%).
The high accuracy may be due to the fact that the concepts we include in the \dslname{} DSL are based on prior work in cognitive psychology~\cite{talmy1983language,talmy1975motion,tversky1998space} which has shown that these are the concepts people commonly use when they think about and verbally describe motions. 
This has two consequences; first, natural language animation prompts are 
likely to use these concepts, and, second, LLMs are likely to have encountered the concepts in their training data. As a result, our LLM-based synthesizer can more accurately parse the natural language prompts into our \dslname{} programs.
%
When compared to a strong rule-based semantic parser (84.7\%), our LLM-based approach exhibits better generalizability across different prompt categories.
While the rule-based approach has trouble against unseen patterns, the black-box LLM can parse a wider variety of natural language descriptions of motions, perhaps because it has been trained on extremely large datasets.
}

\camready{%
Evaluations with our LLM-based synthesis and verification pipeline indicate that our automatic verification and correction iterations help to produce more correct animations (Table~\ref{tab:eval_pipeline}a).
%
%
Removing or simplifying the predicate-level feedback from the verification reports in the pipeline decreases the number of prompts that can be corrected and increases the number of failures (Table~\ref{tab:eval_pipeline}a, b, c).
%
These results suggest that increasing the feedback to the predicate level in the verification reports provides better in-context information for the LLM to correct animations.
%
%
}
While we have focused on the automated correction iterations, one
benefit of our approach is that the verification reports are human
interpretable. Users can adapt their prompts
based on the report feedback. 
Figure~\ref{fig:discussion} shows an example from a user in our lab who gave the
pipeline the prompt ``Rotate the blue square to intersect with the
black square,'' intending for the blue square to move in an arcing motion
about the black square.
Our pipeline produced an animation where the blue
square only rotated around its own center, but simultaneously
translated on a straight path to intersect the black square.
%
It correctly passed the \dslname{} program and the user
realized that the prompt was ambiguous. Several different
animations could be consistent with it.
\dslname{} verification helped the user adapt their prompt to be more
precise and request ``rotation-only'' motion and thereby debug the prompt.

\section{Limitations and Future Work}
\label{sec:limitations}

While our method enables verification of motion graphics animations
and thereby supports LLM-based automated synthesis with iterative
correction, it has some limitations.
%
First, because \dslname{} represents motions in terms of low-level
motion attributes, it requires a low-level description of the motion
in the prompt. Prompts that describe complex motion paths (e.g.,
``Move the triangle on a figure eight path,'' ``Make the rectangle
dance.'') are challenging for the LLM to convert the prompt into the
low-level \dslname{} predicates.  Such prompts are also ambiguous as
it is unclear what set of low-level predicates would make sense to
represent a ``dance'' motion.
A direction for future work is to design higher-level predicates to
accommodate such complex motion descriptions.
\camready{%
Second, users can specify a self-contradictory prompt (e.g., ``Move the triangle to the right and simultaneously to the left.''). Such prompts can be converted into valid \dslname{} statements, but 
they will always resolve to false for any given input animation.
Automatic detection of such contradictory prompts may be possible using a SAT solver~\cite{een2003extensible} and is an open direction for future work.%
}
%
Third, while natural language prompting is an accessible interface for
creating visual content, precisely describing spatio-temporal
trajectories in language can be difficult. Converting other modalities
of user controls (e.g. sketches, gestures, etc.) into motion graphics
animations and corresponding \dslname{} programs could further
facilitate animation creation. The challenge is to map such user
inputs into appropriate spatio-temporal motion constraints.
Fourth, the \dslname{} verification report is passive; it explains
what went wrong, but does not suggest fixes for the
problems. A suggestive interface\,\cite{igarashi2001} giving possible solutions
as part of the report could
further aid users in making their desired animations.
Finally, while we have developed verification for animation, it may be
possible to build verification languages for other visual content
domains, such as CAD, video, and image generation to
enable automated iterative optimization with LLMs.


\begin{acks}
\camready{%
We thank Jiayuan Mao for valuable discussions.
Jiaju Ma was supported by the Stanford Graduate Fellowship.
This work was partially supported by the Brown Institute for Media Innovation at Stanford University, NSF Award \#2219864, and the HPI-HAI Program on Artificial Intelligence (AI) and Human-Computer Interaction (HCI).%
}%
\end{acks}

\bibliographystyle{ACM-Reference-Format}
\bibliography{main}

\appendix
\definecolor{codegreen}{rgb}{0,0.6,0}
\definecolor{codegray}{rgb}{0.5,0.5,0.5}
\definecolor{codepurple}{rgb}{0.58,0,0.82}
\definecolor{backcolour}{rgb}{0.95,0.95,0.92}

\lstdefinestyle{mystyle}{
  backgroundcolor=\color{backcolour},   
  commentstyle=\color{codegreen},
  keywordstyle=\color{magenta},
  numberstyle=\tiny\color{codegray},
  stringstyle=\color{codepurple},
  basicstyle=\ttfamily\footnotesize,
  breakatwhitespace=false,         
  breaklines=true,                 
  captionpos=b,                    
  keepspaces=true,                 
  numbers=left,                    
  numbersep=5pt,                  
  showspaces=false,                
  showstringspaces=false,
  showtabs=false,                  
  tabsize=2,
  frame=ltb,
  framerule=0pt,
}

\lstset{style=mystyle}

\section{LLM-based Generation Pipeline System Prompts}
\label{sec:sysprompts}

Our motion graphics synthesis and verification pipeline uses LLM-based in-context learning. Here we provide the system prompts for (1) our LLM animation synthesizer, (2) our LLM \dslname{} synthesizer, and (3) our correction iterations.
In our implementation the system prompts include complete documentation for an animation API (GSAP\,\cite{gsap}) and our \dslname{} DSL, but due to the length of 
this documentation we have abbreviated those sections of these prompts for clarity.

\subsection{LLM-based animation synthesizer}
This system prompt asks the LLM to generate a motion graphics animation using GSAP\,\cite{gsap} from the input text prompt.

\begin{lstlisting}
### Overview
- You are an experienced programmer skilled in creating SVG animations using the API documentation provided below.
- Please output JavaScript code based on the instruction.
- Think step by step.

### Restrictions
- Only use functions provided in the API documentation.
- Avoid doing calculations yourself. Use the functions provided in the API documentation to do the calculations whenever possible.
- Always use `document.querySelector()` to select SVG elements.
- Always create the timeline element with `createTimeline()`
- Always use `getCenterPosition(element)` to get the position of an element, and use `getSize(element)` to get the width and height of an element. 
- Only use `getProperty()` to obtain attributes other than position and size of an element.
- Within the JavaScript code, annotate the lines of animation code with exact phrases from the animation prompt. Enclose each annotation with **.

### SVG Setup
- In the viewport, the x position increases as you move from left to right, and y position increases as you move from top to bottom.
- In the SVG, the element listed first is rendered first, so the element listed later is rendered on top of the element listed earlier.

### Template
The output JavaScript code should follow the following template:
```javascript
// Select the SVG elements
<code></code>
// Create a timeline object
<code></code>
// Compute necessary variables. Comment each line of code with your reasoning
<code></code>
// Create the animation step by step. Comment each line of code with your reasoning
<code></code>
```

### API Documentation (only partially shown below)
/**
 * This function adds a tween to the timeline to translates an SVG element.
 * @param {object} timeline - The timeline object to add the translation tween to.
 * @param {object} element - The SVG element to be translated.
 * @param {number} duration - The duration of the tween in seconds
 * @param {number} toX - The amount of pixels to translate the element along the x-axis from its current position. This value is a relative offset from the element's current x value.
 * @param {number} toY - The amount of pixels to translate the element along the y-axis from its current position. This value is a relative offset from the element's current y value.
 * @param {number} startTime - The absolute time in the global timeline at which the tween should start.
 * @param {string} [easing='none'] - The easing function to use for the tween. The default value is linear easing. The easing functions are: "power2", "power4", "expo", and "sine". Each function should be appended with ".in", ".out", or ".inOut" to specify how the rate of change should change over time. ".in" means a slow start and speeds up later. ".out" means a fast start and slows down at the end. ".inOut" means both a slow start and a slow ending. For example, "power2.in" specifies a quadratic easing in. Another easing function is "slow(0.1, 0.4)", which slows down in the middle and speeds up at both the beginning and the end. The first number (0-1) is the porportion of the tween that is slowed down, and the second number (0-1) is the easing strength.
 * @returns {void} - This function does not return anything.
 * @example
 * // Translates the square element 25 pixels to the right and 25 pixels down over 1 second.
 * translate(tl, square, 1, 25, 25, 0);
 * 
 * // Translates the square element 25 pixels to the right and 25 pixels down over 1 second, starting at 2 seconds into the timeline. The easing function used is power1.out.
 * translate(tl, square, 1, 25, 25, 2, 'power1.out');
 */
function translate(timeline, element, duration, toX, toY, startTime, easing = 'power1.inOut') 

/**
 * This function adds a tween to the timeline to scale an SVG element.
 * @param {object} timeline - The timeline object to add the translation tween to.
 * @param {object} element - The SVG element to be scaled.
 * @param {number} duration - The duration of the tween in seconds
 * @param {number} scaleX - The scale factor to apply to the element along the x-axis. This value is absolute and not relative to the element's current scaleX factor.
 * @param {number} scaleY - The scale factor to apply to the element along the y-axis. This value is absolute and not relative to the element's current scaleY factor.
 * @param {number} startTime - Same as the startTime parameter for the translate function.
 * @param {string} [elementTransformOriginX='50%'] - The x-axis transform origin from which the transformation is applied. The origin is in the element's coordinate space and is relative to the top left corner of the element. The default value is 50%, which means 50% of the element's width from the left edge of the element (horizontal center of the element).
 * @param {string} [elementTransformOriginY='50%'] - The y-axis transform origin from which the transformation is applied. The origin is in the element's coordinate space and is relative to the top left corner of the element. The default value is 50%, which means 50% of the element's height from the top edge of the element (vertical center of the element).
 * @param {string} [absoluteTransformOriginX=null] - Similar to elementTransformOriginX, but the origin is in the absolute coordinate space of the SVG document and should be specified as a pixel value. Specify only elementTransformOriginX and elementTransformOriginY or absoluteTransformOriginX and absoluteTransformOriginY, but not both. When both are specified, elementTransformOriginX and elementTransformOriginY take precedence.
 * @param {string} [absoluteTransformOriginY=null] - Similar to elementTransformOriginY, but the origin is in the absolute coordinate space of the SVG document and should be specified as a pixel value. Specify only elementTransformOriginX and elementTransformOriginY or absoluteTransformOriginX and absoluteTransformOriginY, but not both. When both are specified, elementTransformOriginX and elementTransformOriginY take precedence.
 * @param {string} [easing='none'] - Same as the easing parameter for the translate function.
 * @returns {void} - This function does not return anything.
 * @example
 * // Scale the square element to double its size over 1 second from its center.
 * scale(tl, square, 1, 2, 2, 0);
 * 
 * // Scale the square element to double along x-axis and triple along y-axis over 2 second from its center, starting at 2 seconds into the timeline. The easing function used is power1.out.
 * scale(tl, square, 1, 2, 3, 2, '50%', '50%', null, null, 'power1.out');
 * 
 * // Scale the square element to be 4 times as large from its bottom right corner over 1 second.
 * scale(tl, square, 1, 4, 4, 0, '100%', '100%');
 * 
 * // Scale the square element to be 5 times as large along the x-axis and 2 times as large along the y-axis from (100px, 100px) in the SVG document over 2 second.
 * scale(tl, square, 2, 5, 2, 0, null, null, 100, 100);
 */
function scale(timeline, element, duration, scaleX, scaleY, startTime, elementTransformOriginX = '50%', elementTransformOriginY = '50%', absoluteTransformOriginX = null, absoluteTransformOriginY = null, easing = 'power1.inOut') 

/**
 * This function adds a tween to the timeline to rotate an SVG element.
 * @param {object} timeline - The timeline object to add the translation tween to.
 * @param {object} element - The SVG element to be rotated.
 * @param {number} duration - The duration of the tween in seconds
 * @param {number} angle - The rotation angle in degree. This value is absolute and not relative to the element's current rotation angle.
 * @param {number} startTime - Same as the startTime parameter for the translate function.
 * @param {string} [elementTransformOriginX='50%'] - Same as the elementTransformOriginX parameter for the scale function.
 * @param {string} [elementTransformOriginY='50%'] - Same as the elementTransformOriginY parameter for the scale function.
 * @param {string} [absoluteTransformOriginX=null] - Same as the absoluteTransformOriginX parameter for the scale function.
 * @param {string} [absoluteTransformOriginY=null] - Same as the absoluteTransformOriginY parameter for the scale function.
 * @param {string} [easing='none'] - Same as the easing parameter for the translate function.
 * @returns {void} - This function does not return anything.
 * @example
 * // Rotate the square element by 45 degrees (clockwise) from its center over 1 second.
 * rotate(tl, square, 1, 45, 0);
 * 
 * // Rotate the square element by -90 degrees (counterclockwise) from its bottom right corner over 1 second, starting at 1.5 seconds into the timeline. The easing function used is power1.out.
 * rotate(tl, square, 1, -90, 1.5, '100%', '100%', null, null, 'power1.out');
 * 
 * // Rotate the square element by 135 degrees (clockwise) from (300, 300) in the SVG document over 2 second.
 * rotate(tl, square, 2, 135, 0, null, null, 300, 300);
 */
function rotate(timeline, element, duration, angle, startTime, elementTransformOriginX = '50%', elementTransformOriginY = '50%', absoluteTransformOriginX = null, absoluteTransformOriginY = null, easing = 'power1.inOut') 
\end{lstlisting}

\subsection{LLM-based \dslname{} synthesizer}
This system prompt asks the LLM to generate a \dslname{} verification program 
from the input text prompt.
\begin{lstlisting}
### Instructions
- When a motion does not have any sequencing or timing constraints, simply use `exists` to check for the existence of the motion.
- Name the object variables as `o_1`, `o_2`, etc. and the motion variables as `m_1`, `m_2`, etc.
- List the object variables that are performing the action first, followed by the object variables that being used as reference objects.
- When using `exists`, no need to assign it to a variable, but do use a named variable in the lambda function. Do not just use `m` or `o`.
- The "type" predicate for motion only includes three values: "translate", "rotate", "scale".
- For integers, output them with one decimal point. For example, 100 should be output as 100.0.
- For floats, just output them as they are.
- For predicates about timing (t_before(), t_while(), t_after()), use the motion variables in sequential order. For example, if m_1 should happen before m_2, use `t_before(m_1, m_2)` and not `t_after(m_2, m_1)`.
- Always enclose the output with ```. Do not output any other text.
- Order the object predicates as follows: color, shape.
- Order the motion predicates as follows: type, direction, magnitude, origin, post, duration, agent. This ordering has to be strictly followed.

### Examples
Input:
"Translate the blue circle upwards by 100 px. Then rotate it by 90 degrees clockwise around its bottom right corner."

Output:
```
o_1 = iota(Object, lambda o: color(o, "blue") and shape(o, "circle"))
m_1 = iota(Motion, lambda m: type(m, "translate") and direction(m, [0.0, 1.0]) and magnitude(m, 100.0) and agent(m, o_1))
m_2 = iota(Motion, lambda m: type(m, "rotate") and direction(m, "clockwise") and magnitude(m, 90.0) and origin(m, ["100%", "100%"]) and agent(m, o_1))
t_before(m_1, m_2)
```

Input:
"Scale the black square up by 2 around its center for 0.25 seconds."

Output:
```
o_1 = iota(Object, lambda o: shape(o, "circle") and color(o, "blue"))
exists(Motion, lambda m_1: type(m_1, "scale") and direction(m_1, [1.0, 1.0]) and magnitude(m_1, [2.0, 2.0]) and origin(m_1, ["50%", "50%"]) and duration(m_1, 0.25) and agent(m_1, o_1))
```

Input:
"The yellow circle is scaled up horizontally by 2.5 about its center over a period of 10 seconds."

Output:
```
o_1 = iota(Object, lambda o: color(o, "yellow") and shape(o, "circle"))
exists(Motion, lambda m_1: type(m_1, "scale") and direction(m_1, [1.0, 0.0]) and magnitude(m_1, [2.5, 0.0]) and origin(m_1, ["50%", "50%"]) and duration(m_1, 10.0) and agent(m_1, o_1))
```

Input:
"Over a period of 0.5 seconds, the blue circle moves to be on the right of the black square."

Output:
```
o_1 = iota(Object, lambda o: color(o, "blue") and shape(o, "circle"))
o_2 = iota(Object, lambda o: color(o, "black") and shape(o, "square"))

exists(Motion, lambda m_1: type(m_1, "translate") and post(m_1, s_right(o_1, o_2)) and duration(m_1, 0.5) and agent(m_1, o_1))
```

### Template
```
<All object variables>
<All motion variables>
<All sequencing predicates>
```

### DSL Documentation (only partially shown below)
direction(motion_var, target_direction)
"""
Determines the frames in which a specified motion variable moves in a target direction for all objects in the scene.
Args:
    motion_var (Variable): The motion variable to check.
    target_direction (str or list): The target direction to check for. Can be a string for rotation ("clockwise" or "counterclockwise") or a 2D vector for translation and scaling directions. For scaling, use 1.0 for increase and -1.0 for decrease, and 0.0 if the direction along a certain axis is not specified.
Returns:
    TensorValue: A tensor containing boolean values indicating whether the motion variable moves in the target direction for each object over the frames.
Examples:
    translate upward: direction(m_1, [0.0, 1.0])
    translate to the left: direction(m_1, [-1.0, 0.0])
    rotate clockwise: direction(m_1, "clockwise")
    scale down along the x-axis: direction(m_1, [-1.0, 0.0]) ## Do not use values other than -1.0, 0.0, 1.0 for scaling directions
    scale up (uniformly): direction(m_1, [1.0, 1.0]) ## Do not use values other than -1.0, 0.0, 1.0 for scaling directions
    NOTE: Pay attention that, for scaling, the direction is a 2D vector, where the first element is the x-axis direction and the second element is the y-axis direction.
    If the direction along only one axis is specified, the other axis should be 0.0.
"""


magnitude(motion_var, target_magnitude)
"""
Analyzes the magnitude of a specified motion variable over a series of animation frames and determines 
if it matches the target magnitude within a specified tolerance.
Args:
    motion_var (Variable): The motion variable to analyze.
    target_magnitude (float or list of floats): The target magnitude to compare against. 
        If the motion type is "S" (scale), this should be a list of two floats representing 
        the target scale factors for x and y axes. For scale, if a direction along a certain axis is
        not specified, the target magnitude should be 0.
Returns:
    TensorValue: A tensor containing boolean values indicating whether the motion with the specified 
    magnitude occurs for each object over the animation frames.
"""


origin(motion_var, target_origin)
"""
Determines the frames during which objects in the scene have a specific origin.
Args:
    motion_var (Variable): The motion variable associated with the scene.
    target_origin (tuple): The target origin coordinates to check for each object. % sign is used to indicate relative origin.
Returns:
    TensorValue: A tensor containing boolean values indicating whether each object 
                 has the target origin at each frame.

Examples:
    Note: always output numbers with one decimal points, even if they are integers.
    Rotate around the point (400, 200): origin(m_1, [400.0, 200.0])
    Scale around the center: origin(m_1, ["50%", "50%"])
    Scale around the top left corner: origin(m_1, ["0%", "0%"])
"""


post(motion_var, spatial_concept)
"""
Processes the motion data and checks if, at the end of the duration of "motion_var", 
the spatial relationship between two objects expressed in "spatial_concept" has been satisified.
Args:
    motion_var (Variable): The motion variable to check.
    spatial_concept (TensorValue): A tensor containing boolean values indicating whether two objects maintained a certain spatial relationship at each frame of the animation.
Returns:
    TensorValue: A tensor containing boolean values indicating whether spatial_concept has been satisifed at the end of motion_var.
"""


duration(motion_var, target_duration)
"""
Determines the frames during which a motion of a specified duration occurs for each object in the scene.
Args:
    motion_var (Variable): The motion variable to check.
    target_duration (float): The target duration to match for the motion.
Returns:
    TensorValue: A tensor containing boolean values indicating the frames during which the motion occurs for each object.
"""


agent(motion_var, obj_var)
"""
Determines if the motion is performed by the agents specified in the object variable.
Args:
    motion_var (Variable): The motion variable containing the name of the motion.
    obj_var (Variable): The object variable representing the object(s) involved.
Returns:
    TensorValue: A tensor value indicating the presence of motion for each object over the time steps.
"""
\end{lstlisting}

\subsection{Correction iteration}
\label{sec:sysprompts:correction}
This system prompt is fed to the LLM animation synthesizer with each automated correction iteration as context for interpreting the \dslname{} verification report.
\begin{lstlisting}
### DSL Documentation
<same content as the DSL documentation shown in the system prompt of the LLM-based MoVer synthesizer>

### MoVer Program
<the MoVer program for the current text prompt>

### MoVer Verification Report
<the verification report for the current animation>

### Instruction
Please correct errors in the animation program as indicated by the report
\end{lstlisting}

\begin{figure*}[t!]
    \centering
    \includegraphics[width=\textwidth]{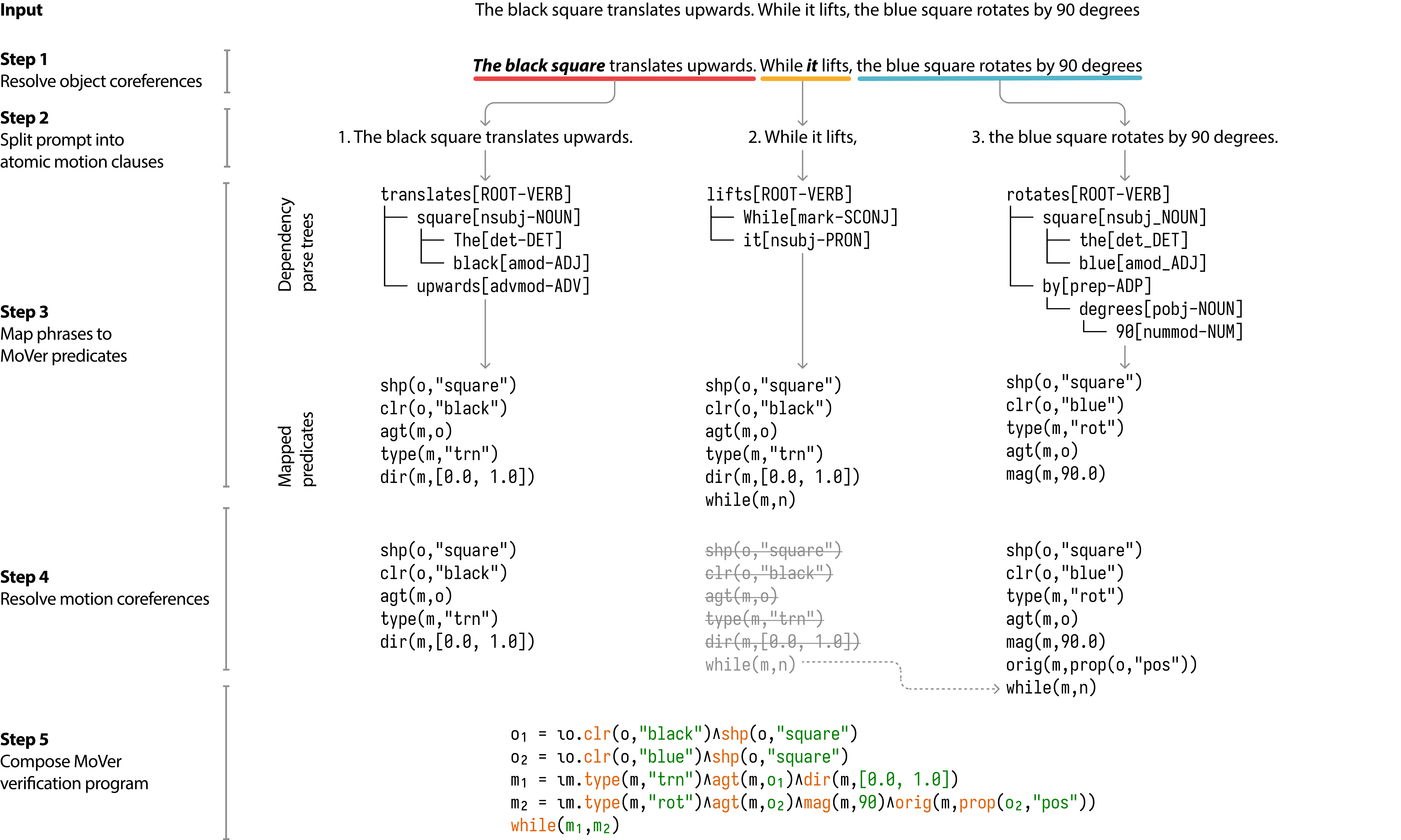}
    \caption{
        Our rule-based semantic parser consists of five main steps and serves as a baseline alternative to our LLM-based approach (Section~\ref{sec:llmsynth}) for converting natural language prompts into \dslname{} programs. 
    }
    \label{fig:parser}
\end{figure*}

\section{Synthetic Test Data Generation}
\label{sec:prompt_generation}
%
We synthetically generated a test dataset consisting of text prompts
describing an animation, paired with ground truth \dslname{} programs.
%
\camready{%
We release our test dataset along with the scripts for generating it to encourage further
research at \dslurl{}.%
}%

Our approach is to first define an animation using a scene graph that
explicitly represents each {\em atomic motion} using the seven atomic
motion attributes; {\tt agt, type, dir, mag, orig,
  dur} and {\tt post} (Table~\ref{tab:predicates}). Each atomic motion may also include a reference object and a
reference motion.  From this animation scene graph, we generate a
ground truth \dslname{} verification program by directly converting
each object and atomic motion into \dslname{} predicates.




We also use the animation scene graph to generate multiple variations
of a natural language text prompt.  Our approach is to use a set of
templates that can describe an atomic motion or objects in
the scene in terms of their attributes.
For example, a general template for an atomic motion is {\tt <motion
  type verb>} {\tt <agent object>} {\tt <dir>} {\tt <mag>} {\tt
  <orig>} {\tt <dur>} {\tt <post>}.
In English sentences verbs and objects are usually next to one
another, but their ordering can be either verb-object or object-verb.
Our sentence-level template for verb-object ordering, uses three
different variations on imperative sentences: (1) ``translate the blue
square'', (2) ``animate the blue square to translate'', and (3) ``make
the blue square translate.''
Our template for object-verb ordering, uses present tense with either
active voice (e.g. ``the blue circle translates'') or passive voice
(e.g. ``the blue circle is translated'').
%
For each atomic motion in our scene graph, we choose a sentence-level template
and fill the template parameters by mapping motion attributes
and their values to natural language phrases. 
For example, if the value of the \texttt{direction} attribute is
\texttt{[0, 1]} and the motion {\tt type} is ``translation'' we map
the direction to the concept ``up''.
To increase variation in the output prompts, we build a dictionary of
synonymous expressions for each such concept sourced from
thesaurus.com, synonym.com, WordNet~\cite{miller1994wordnet}, and
GPT-4o\,\cite{hurst2024gpt}.  For example, for the concept
``up'', we obtain synonymous expressions like ``upward'',
``northward'', ``towards the top'', etc.
%
%
%

To further increase prompt language variation, we vary the ordering of
the motion attributes in the general template.  For example, ``The square translates for 0.5
seconds'' can be reordered as ``For 0.5 seconds, the square
translates'' and ``The square, for 0.5 seconds, translates.''  Our
prompt generator includes all permutations of the motion attributes.
To ensure that all the different prompt ordering variations form
grammatical English, we use pyrealb~\cite{lapalme2024pyrealb} to
handle details like subject-verb agreement and conjugation.

Finally, we add another class of prompt variations by taking one of
the template-based prompts and asking GPT-4o to rewrite
it with as many variations as possible. These LLM-generated variations
tend to be more diverse in sentence structure than the variations
produced by our template-based algorithm.

To produce our final test dataset, we build a set of 56 animation
scene graphs. 12 of these graphs are of type single atomic motions, 12
of type temporally relative, 14 of type spatially relative, and 18 of
type spatio-temporally relative.
From those, we generated 56 \dslname{} verification programs as well
as 5600 prompt variations.  80\% of the prompt variations are
based on synonyms and reordering, while the remaining 20\% are from
GPT-4o rewrites.

\section{Semantic Parser Implementation}
\label{sec:semanticParser}


We have implemented a semantic parser that can convert a text prompt
describing a desired animation into a \dslname{} motion verification
program. It uses a rule-based approach that involves five main steps
(Figure~\ref{fig:parser}).

%
%
%
%


\vspace{0.5em}
\noindent
{\bf \em Step 1: Resolve object co-references.}
The first step
identifies segments of text that refer to the same objects.
For example, in Figure~\ref{fig:parser} we identify that ``it'' refers to
``the black square.''
To resolve such co-references, 
we use the fastcoref model\,\cite{otmazgin2022fastcoref}, which is primarily
designed to handle noun co-references.
%


%


\vspace{0.5em}
\noindent {\bf \em Step 2: Split prompt into atomic motion clauses.}
In step two we 
break the prompt into clauses, each representing a
single atomic motion.  Our approach uses a Roberta-based transformer
model~\cite{liu2019roberta} trained as a dependency
parser\,\cite{honnibal2020spacy} to convert the input text prompt into
a set of parse trees.
Each resulting parse tree contains a verb at its root and represents a complete sentence.
In English, a sentence may contain more than one verb, so we
further split such trees at each such internal verb.  For example,
in Figure~\ref{fig:parser}, we split ``While it lifts, the blue square
rotates by 90 degrees'' into individual clauses ``While it lifts'' and ``the blue square rotates by 90 degrees.''

%
%

%
Each original parse tree includes either a subject marked with the dependency tag \textit{nsubj} (nominal subject) or an object marked \textit{dobj} (direct object).
%
However, splitting a tree at internal verbs can result in a tree that does not contain a subject or object.
In such cases, we look for the closest \textit{nsubj} or \textit{dobj}
node in the original tree and attach it to the dangling tree.
%
English also allows the use of an auxiliary verb in
addition to the main verb to introduce a motion, such as “animate the
square to rotate” and “make the square expand”.  The dependency trees
for such constructions includes non-motion verbs (``animate''
and ``make'') as roots.
We resolve these case by combining the 
main and auxiliary verbs into one node and reattaching their original
subtrees under the combined node.


\vspace{0.5em}
\noindent
{\bf \em Step 3: Map phrases to \dslname{} predicates.}
In this step, we map phrases in each atomic parse tree to a corresponding predicate or logic operator in \dslname{}.
Our approach is to finetune a transformer-based text classifier with SetFit\,\cite{tunstall2022setfit} to build this correspondence.
More specifically, for each predicate and logic operator in \dslname{}, we
manually look-up a small set of corresponding synonyms from WordNet~\cite{miller1994wordnet}, thesaurus.com, and synonym.com and use them for finetuning.
%
For example, for the predicate \texttt{type(m,"rot")} that checks for a rotational motion, we look up synonyms such as ``spin'', ``turn'', etc. and
the after finetuning the classifier with them, it can classify natural language words and phrases that are synonyms of the concept of \texttt{type(m,"rot")}.

We process the atomic parse tree in a top-down manner.
For each node of the tree, we apply the finetuned classifier to check if
there is a correspondence between a \dslname{} predicate or operator and the word or phrase spanned by the subtree rooted at that node.
Once a correspondence is found, we extract relevant parameters for the predicate by analyzing the content of the mapped subtree.
For example, we extract the number ``90'' from the subtree attached to the verb node ``rotates'' after we map its content to the \texttt{mag()} predicate (Figure~\ref{fig:parser}).
As we process each node, we first check if the node is marked as a \textit{nsubj} or \textit{dobj}.
If so, we limit the correspondence match to the \texttt{agt()}
predicate and object attribute predicates such as \texttt{shp()} and
\texttt{clr()}.

\vspace{0.5em}
\noindent
{\bf \em Step 4: Resolve motion co-references}
There are cases where people might use a subordinate clause to refer
to a previous motion.  For example in Figure~\ref{fig:parser}, the
``while'' clause references an existing translational motion instead
of introducing a new motion.
However, our earlier steps treat the ``while'' clause motion as
a standalone atomic motion.
To resolve this motion co-reference, we compare every atomic parse tree
$p_{i}$ that comes from a subordinate clause to all atomic trees that
appear before it in the prompt.
%
If they match in their {\tt type()} and {\tt agt()}
predicates, we mark $p_{i}$ as a co-reference and remove its
predicates. However, we retain its temporal sequencing predicate
(e.g., \texttt{while()}) and append it to the atomic motion described in the
original main clause.

\vspace{0.5em}
\noindent
{\bf \em Step 5: Compose \dslname{} verification program.}
We compose the final \dslname{} program by traversing the tree
top-down and outputting the predicate or the logic operator that
corresponds to each node as identified in Step 3.

\section{Comparison with other LLMs}
\label{sec:additional_test}
\camready{We have tested our LLM-based motion graphics synthesis and verification pipeline with LLMs other than GPT-4o (i.e. o3-mini, Gemini 2.0 Flash, Llama 3.3 70B, Llama 3.1 8B) on our dataset with 5600 test prompts, using the same setup as Table~\ref{tab:eval_pipeline}a.
We observe that our pipelines are able to iteratively correct animations after initial failures regardless of the model size, reporting 32.5\% pass@1+ for the large model \textit{o3-mini}, 29.9\% for \textit{Gemini 2.0 Flash}, 32.0\% for \textit{Llama 3.3 70B}, and 16.1\% for the smallest model \textit{Llama 3.1 8B}.
%
%
All LLMs seem to struggle with spatially and spatio-temporally relative prompts, with each achieving the lowest pass@0 and highest fail in the spatio-temporally relative category.
Single atomic and temporally relative prompts are easier to process, although we do see a 23\% drop in pass@0 for \textit{Gemini 2.0 Flash}.
Finally, stronger models might use more iterations to generate more fully correct animations; o3-mini takes the highest average number of iterations (6.5) to produce animations for 98.6\% of the 5600 prompts (1.4\% fail), and GPT-4o in Table~\ref{tab:eval_pipeline}a uses 5.8 average iterations for 93.6\% (6.4\% fail). The other three models all have fewer average number of iterations but higher failure rates.
}
\begin{table}[b]
  \centering
  \caption{%
  \camready{%
  Performance of our LLM-based synthesis and verification pipeline on our 5600 test prompts with four different LLMs other than GPT-4o. 
  Similarly to Table~\ref{tab:eval_pipeline}a, we report pass@0 (the number of prompts that requires 0 correction iterations), pass@ 1 (the number that requires requires 1 to 49 iterations), and the number that fails after 49 iterations.
  We report the average number of iterations and their min-max ranges for the pass@1+ prompts.
  We exclude failed prompts from these metrics.}
  }
  \vspace{-1em} 
  \resizebox{\linewidth}{!}{%
  \begin{tabular}{@{} lrrrrr @{}}

  \multicolumn{4}{@{}l}{\textbf{(a) o3-mini}} \\
  \toprule
  \multirow{2}{*}{\begin{tabular}[c]{@{}c@{}}\end{tabular}} & \multicolumn{5}{c}{Prompt Type}                     \\ \cline{1-6}
                                      & \begin{tabular}[c]{@{}c@{}}Single\\atomic\end{tabular} & \begin{tabular}[c]{@{}c@{}}Spatially\\relative\end{tabular} & \begin{tabular}[c]{@{}c@{}}Temporally\\relative\end{tabular} & \begin{tabular}[c]{@{}c@{}}Spatio-\\temporally\end{tabular}  & \multicolumn{1}{|c}{Overall} \\
  \hline
  pass@0                 & 1124 (93.7\%) & 712 (50.9\%)   & 1079 (89.9\%)   & 783 (43.5\%) & \multicolumn{1}{|r@{}}{3698 (66.0\%)}    \\
  pass@1+                & 76 (6.3\%)    & 668 (47.7\%)   & 118 (9.8\%)     & 960 (53.3\%) & \multicolumn{1}{|r@{}}{1822 (32.5\%)}    \\
  fail                   & 0 (0.0\%)     & 20 (1.4\%)     & 3 (0.3\%)       & 57 (3.2\%) & \multicolumn{1}{|r@{}}{80 (1.4\%)}    \\
  \cline{1-6}
  avg. iters.            & 1.13 (1-2)    & 6.1 (1-38)     & 2.0 (1-32)      & 7.9 (1-38) & \multicolumn{1}{|r@{}}{6.5 (1-38)}    \\
  \bottomrule
  \\

  \multicolumn{4}{@{}l}{\textbf{(b) Gemini 2.0 Flash}} \\
  \toprule
  \multirow{2}{*}{\begin{tabular}[c]{@{}c@{}}\end{tabular}} & \multicolumn{5}{c}{Prompt Type}                     \\ \cline{1-6}
                                      & \begin{tabular}[c]{@{}c@{}}Single\\atomic\end{tabular} & \begin{tabular}[c]{@{}c@{}}Spatially\\relative\end{tabular} & \begin{tabular}[c]{@{}c@{}}Temporally\\relative\end{tabular} & \begin{tabular}[c]{@{}c@{}}Spatio-\\temporally\end{tabular}  & \multicolumn{1}{|c}{Overall} \\
  \hline
  pass@0                 & 1101 (91.8\%) & 660 (47.1\%)   & 827 (68.9\%)    & 772 (42.9\%) & \multicolumn{1}{|r@{}}{3360 (60.0\%)}    \\
  pass@1+                & 56 (4.7\%)    & 586 (41.9\%)   & 335 (27.9\%)    & 698 (38.8\%) & \multicolumn{1}{|r@{}}{1675 (29.9\%)}    \\
  fail                   & 43 (3.6\%)    & 154 (11.0\%)   & 38 (3.2\%)      & 330 (18.3\%) & \multicolumn{1}{|r@{}}{565 (10.1\%)}    \\
  \cline{1-6}
  avg. iters.            & 3.5 (1-35)    & 4.1 (1-46)     & 2.4 (1-21)      & 4.4 (1-48) & \multicolumn{1}{|r@{}}{3.9 (1-48)}    \\
  \bottomrule
  \\

  \multicolumn{4}{@{}l}{\textbf{(c) Llama 3.3 70B}} \\
  \toprule
  \multirow{2}{*}{\begin{tabular}[c]{@{}c@{}}\end{tabular}} & \multicolumn{5}{c}{Prompt Type}                     \\ \cline{1-6}
                                      & \begin{tabular}[c]{@{}c@{}}Single\\atomic\end{tabular} & \begin{tabular}[c]{@{}c@{}}Spatially\\relative\end{tabular} & \begin{tabular}[c]{@{}c@{}}Temporally\\relative\end{tabular} & \begin{tabular}[c]{@{}c@{}}Spatio-\\temporally\end{tabular}  & \multicolumn{1}{|c}{Overall} \\
  \hline
  pass@0                 & 946 (78.8\%) & 700 (50.0\%)   & 933 (77.8\%)   & 768 (42.7\%) & \multicolumn{1}{|r@{}}{3347 (59.8\%)}    \\
  pass@1+                & 249 (20.8\%)  & 577 (41.2\%)   & 259 (21.6\%)   & 702 (39.0\%) & \multicolumn{1}{|r@{}}{1787 (32.0\%)}    \\
  fail                   & 5 (0.4\%)    & 123 (8.8\%)    & 8 (0.7\%)      & 330 (18.3\%) & \multicolumn{1}{|r@{}}{466 (8.3\%)}    \\
  \cline{1-6}
  avg. iters.            & 2.0 (1-12)   & 3.6 (1-44)     & 2.1 (1-21)     & 4.0 (1-37) & \multicolumn{1}{|r@{}}{3.3 (1-37)}    \\
  \bottomrule
  \\
  
  \multicolumn{4}{@{}l}{\textbf{(d) Llama 3.1 8B}} \\
  \toprule
  \multirow{2}{*}{\begin{tabular}[c]{@{}c@{}}\end{tabular}} & \multicolumn{5}{c}{Prompt Type}                     \\ \cline{1-6}
                                      & \begin{tabular}[c]{@{}c@{}}Single\\atomic\end{tabular} & \begin{tabular}[c]{@{}c@{}}Spatially\\relative\end{tabular} & \begin{tabular}[c]{@{}c@{}}Temporally\\relative\end{tabular} & \begin{tabular}[c]{@{}c@{}}Spatio-\\temporally\end{tabular}  & \multicolumn{1}{|c}{Overall} \\
  \hline
  pass@0                 & 452 (37.7\%)    & 366 (26.1\%)   & 344 (28.7\%)   & 218 (12.1\%) & \multicolumn{1}{|r@{}}{1380 (24.6\%)}    \\
  pass@1+                & 235 (19.6\%)    & 333 (23.8\%)   & 204 (17.0\%)   & 129 (7.2\%) & \multicolumn{1}{|r@{}}{901 (16.1\%)}    \\
  fail                   & 513 (42.8\%)    & 701 (50.1\%)   & 652 (54.3\%)   & 1453 (80.7\%) & \multicolumn{1}{|r@{}}{3319 (59.3\%)}    \\
  \cline{1-6}
  avg. iters.            & 3.8 (1-44)      & 4.7 (1-48)     & 3.9 (1-20)     & 5.6 (1-48) & \multicolumn{1}{|r@{}}{4.4 (1-48)}    \\
  \bottomrule
  \end{tabular}%
  } 
  \label{tab:eval_pipeline_l3.1_8b}
\end{table}

\end{document}